\documentclass[reprint,showpacs,twocolumn,superscriptaddress,aps,prl]{revtex4-1}
\usepackage[utf8]{inputenc}
\usepackage[american,british]{babel}
\usepackage[T1]{fontenc}
\usepackage[pdftex]{graphicx}  
\usepackage{graphicx, xcolor}
\usepackage{dcolumn}
\usepackage{braket}
\usepackage{bm}
\usepackage{amsmath,amsthm,amssymb}
\usepackage{color}
\usepackage{verbatim}
\usepackage{ulem}

\definecolor{darkGreen}{RGB}{0,110,0}
\definecolor{darkBlue}{RGB}{0,0,130}
\usepackage{hyperref}

\hypersetup{
 colorlinks=true,
 linkcolor=blue,
 anchorcolor = blue,
 citecolor = blue,
 filecolor = blue,
 urlcolor = blue
}

\newcommand{\ee}{\mathrm{e}}
\usepackage{dsfont}
\usepackage{physics}

\newcommand{\beginsupplement}{
\clearpage
\pagebreak
\setcounter{equation}{0}
\setcounter{figure}{0}
\setcounter{table}{0}
\setcounter{page}{1}
\makeatletter
\renewcommand{\theequation}{S\arabic{equation}}
\renewcommand{\thefigure}{S\arabic{figure}}
\onecolumngrid
\renewcommand{\thesection}{S\arabic{section}}
\section*{\large{Supplemental Material}}
}

\begin{document}
\title{The Negativity Hamiltonian: An operator characterization of mixed-state entanglement}
\date{\today}

 \author{Sara Murciano}
 \affiliation{SISSA, via Bonomea 265, 34136 Trieste, Italy}
 \affiliation{INFN Sezione di Trieste, via Bonomea 265, 34136 Trieste, Italy}

 \author{Vittorio Vitale}
 \affiliation{SISSA, via Bonomea 265, 34136 Trieste, Italy}
 \affiliation{The Abdus Salam International Center for Theoretical Physics, Strada Costiera 11, 34151 Trieste, Italy}

 \author{Marcello Dalmonte}
 \affiliation{SISSA, via Bonomea 265, 34136 Trieste, Italy}
 \affiliation{The Abdus Salam International Center for Theoretical Physics, Strada Costiera 11, 34151 Trieste, Italy}

 \author{Pasquale Calabrese}
 \affiliation{SISSA, via Bonomea 265, 34136 Trieste, Italy}
 \affiliation{INFN Sezione di Trieste, via Bonomea 265, 34136 Trieste, Italy}
 \affiliation{The Abdus Salam International Center for Theoretical Physics, Strada Costiera 11, 34151 Trieste, Italy}

\begin{abstract}
In the context of ground states of quantum many-body systems, the locality of entanglement between connected regions of space is directly tied to the locality of the corresponding entanglement Hamiltonian: the latter is dominated by local, few-body terms. 
In this work, we introduce the negativity Hamiltonian as the (non hermitian) effective Hamiltonian operator describing the logarithm of the partial transpose of a many-body system. 
This allows us to address the connection between entanglement and operator locality beyond the paradigm of bipartite pure systems. As a first step in this direction, we study the structure of the negativity Hamiltonian for fermionic conformal field theories and a free fermion chain: in both cases, we show that the negativity Hamiltonian assumes a quasi-local functional form, that is captured by simple functional relations. 
\end{abstract} 

\maketitle
\paragraph{\it Introduction -} Over the past two decades, entanglement has been a central concept in many
branches of quantum physics ranging from quantum information~\cite{nielsen2002quantum,benenti2004principles} to condensed matter theory~\cite{amico2002,LAFLORENCIE20161} and high-energy physics~\cite{SrednickiPRL1993,Dong2016,RT,Raamsdonk,maldacena}. In particular, it has been successfully utilized to characterize quantum many-body systems both theoretically and experimentally ~\cite{Brydges2019probing,vitale2021symmetryresolved,elben2020mixed,neven2021symmetryresolved,exp-2016,exp-2019,islam}. The main object which enters in its quantification is the reduced density matrix (RDM). For a given state $\rho$, the RDM of a region $A$, $\rho_A$, is obtained by tracing $\rho$ over the complement of $A$, $B$, that is:
\begin{equation}
   \rho_A=\mathrm{Tr}_B \rho=\frac{e^{-H_A}}{Z_A}, \quad Z_A=\mathrm{Tr}e^{-H_A},
\end{equation}
where the operator $H_A$ is the entanglement (or modular) Hamiltonian (EH).

From a many-body viewpoint, the entanglement properties of pure states can be construed in a hierarchical manner. Firstly, there exists a characterization of its entanglement properties via entanglement entropies. Those are uniquely dependent on the spectrum of $H_A$ - also known as entanglement spectrum. Secondly, it is possible to characterize the properties of the RDM directly at the operator level, via the full characterization of the EH - a paradigmatic example being the Li-Haldane conjecture in the context of topological matter \cite{Haldane}.

The EH fully characterizes the ``local'' properties of entanglement in a many-body system - that is, it allows to understand whether the RDM can be interpreted as the exponential of a local operator composed solely of few-body local terms. In the context of quantum field theory, this principle of locality is an established pillar - the Bisognano-Wichmann theorem~\cite{Bisognanoquantum,Arias2017}. Such locality is at the heart of several physical phenomena - from topological order, to the nature of area laws in gapped systems-, and is the key element at the basis of theory and experiments aimed at large-scale reconstructions of the RDM \cite{Kokail2021,Kokail2021nature,Zhu2019}. However, it is presently unknown whether it is possible to associate locality and entanglement in a similar way for the case of mixed-state entanglement, that encompasses a variety of scenarios of key experimental and theoretical relevance - from mixed states, to correlations between partitions in pure states.
\begin{figure}[b]
    \centering
    \includegraphics[width=1.\linewidth]{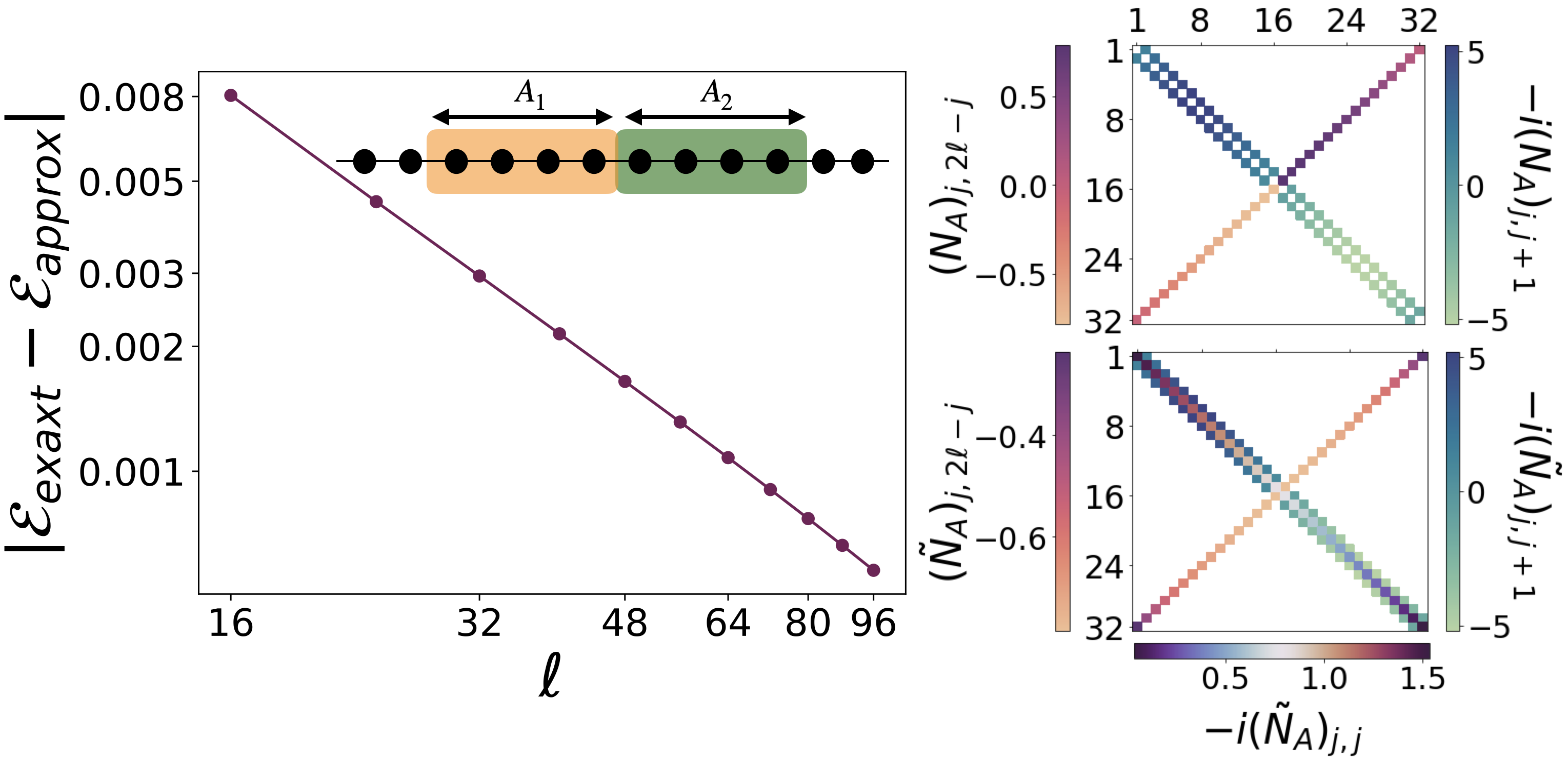}
    \caption{Summary of our results for adjacent intervals of equal length $\ell$ on the infinite line for lattice free fermions (geometry in the inset). The one-particle negativity hamiltonians $N_A$ and $\tilde{N}_A$ are dominated by quasi-local terms appearing close to the diagonal and on the antidiagonal (see the right panels for $\ell=8$).
    Left panel: Comparison of the exact logarithmic negativity  with the approximate one coming from field theoretical $\tilde{N}_A$, see text.}
    \label{fig:sketch1}
\end{figure}
In this work, we introduce and investigate the negativity Hamiltonian - an operator that allows us to cast the relation between locality and entanglement (in particular, that related to Peres-Horodecki criterion) for general mixed states. Our work is directly motivated by a series of recent results that have emphasized the importance of the entanglement negativity in a variety of settings, including harmonic oscillator chains \cite{neg-hc1,neg-hc2,neg-hc3,neg-hc5,neg-hc6,neg-hc7,neg-hc8}, 
quantum spin models \cite{neg-qsm,neg-qsm1,neg-qsm,neg-qsm1,neg-qsm2,neg-qsm6,neg-qsm7,neg-qsm8,neg-qsm9,neg-qsm10,PRXQuantum.2.030347}, free-fermionic systems \cite{shapourian2017partial,shapourian2019twisted,ksrs2017,neg-ferm, Eisler_2016,ez-15}, 
(1+1)d conformal and integrable field theories \cite{negf,cct-12,cct-13,ctt-13b,cct-15,rac-2016,Blondeau,ca-2016,ca-19,ca2-19,ares-2021}, out-of-equilibrium settings \cite{Eisler2014,neg-out,ryu-2020,neg-out1,shi2021entanglement,neg-out2,neg-out3,neg-out4,sara2021quench}, and topological order \cite{lu2020detecting,castelnovo-13,castelnovo-18,neg-top,neg-top1,neg-top2}. Importantly, the negativity is directly linked to the partial transpose $\rho_A^{T_1}$ of the RDM - and, as such, does lend itself naturally to an interpretation based on statistical mechanics. For the case of a subpartition of $A=A_1 \cup A_2$, we define the {\it{negativity Hamiltonian}} ${\cal N}_A$ as
\begin{equation}
\rho_A^{T_1}=Z_A^{-1}e^{-{\cal N}_A}.
\label{NAdef}
\end{equation}
Clearly ${\cal N}_A$ is non-hermitian because negative eigenvalues of $\rho_A^{T_1}$ are the signature of mixed-state entanglement.
Nevertheless, it is still natural to wonder about the locality properties of ${\cal N}_A$ and about the location of its eigenvalues in the complex plane.

After discussing the definition of ${\cal N}_A$ for both bosonic (spin) and fermionic systems, we unveil the operator structure of ${\cal N}_A$ for two relevant cases: (1+1)-d fermionic conformal field theory and a tight-binding model of spinless fermions on a chain. Both cases show a characteristic quasi-local (in a sense to be specified below) structure - a first demonstration of the relation between entanglement and locality at the operator level beyond  the case of complementary partitions. On top of its conceptual relevance, 
and similarly to what has been discussed in the context of pure states for the case of local EHs, this fact enables some immediate consequences: i) interpreting the negativity spectrum, i.e. the analog of the pure-state entanglement spectrum for mixed states \cite{rac-2016,PRXQuantum.2.030347}, ii) simulating this object in nowadays available quantum platforms \cite{Brydges2019probing} iii) applying well-established statistical mechanics tools such as tensor networks \cite{SCHOLLWOCK201196,dmrg} and quantum Monte Carlo  \cite{qmc} to access the entire partial transpose $\rho_A^{T_1}$.

\paragraph{\it The partial transpose  -} To introduce the concept of the negativity Hamiltonian, the first step is to discuss the partial transpose for bosonic and fermionic systems.
Let us start considering a bosonic system $A=A_1\cup A_2$ described by \begin{equation}\rho_A=\sum_{i,j,k,l} \bra{e^{A_1}_i,e^{A_2}_j}\rho_A \ket{e^{A_1}_k,e^{A_2}_l} \ket{e^{A_1}_i,e^{A_2}_j} \bra{e^{A_1}_k,e^{A_2}_l},
\end{equation}
where $\ket{e^{A_1}_i}, \ket{e^{A_2}_j}$ denote orthonormal bases in the Hilbert spaces $\mathcal{H}_{A_1}$ and $\mathcal{H}_{A_2}$ corresponding to subsystems $A_1$ and $A_2$.
The partial transpose of the reduced density matrix $\rho_A^{T_{1}}$ with respect to the system $A_1$ is defined performing a standard transposition in $\mathcal{H}_{A_1}$, i.e. exchanging the matrix elements in $A_1$,
$\rho^{T_{1}}_A=(T_{A_1} \otimes \mathds{1}_{A_2})\rho_A$.
The presence of negative eigenvalues of $\rho^{T_{1}}_A$ is a signature of mixed state entanglement \cite{peres1996separability}, which can be quantified by the logarithmic negativity ${\cal E}= \log \Tr |\rho^{T_{1}}_A|$ \cite{vw-02}.

The partial transposition has also an interpretation in terms of a time-reversal transformation or mirror reflection in phase space \cite{Simon2000}. Namely, considering the one-to-one correspondence between density matrices and Wigner distribution functions $W(q,p)$ then $
\rho_A \rightarrow \rho^T_A \Longleftrightarrow W(q,p) \rightarrow W(q,-p)$.
This can be conveniently observed starting from a bosonic density matrix written in a coherent state basis, since time-reversal transformation ($\mathcal{T}$) can be identified with the complex conjugation~\cite{shapourian2017partial}. Taking $\ket{\alpha}$, a bosonic coherent state, one has
\begin{equation}\label{eq:TRcoherentstate}
    (\ket{\alpha}\bra{\alpha^{*}})\overset{\mathcal{T}}{\longrightarrow}\ket{\alpha^{*}}\bra{\alpha}=(\ket{\alpha}\bra{\alpha^{*}})^T.
\end{equation}

In the case of fermionic systems, the equivalence above does not hold and the definition of partial transposition differs when looking at the density matrix or at the Wigner distribution function.
In a coherent state basis the RDM reads \cite{ez-15,shapourian2017partial,ShapourianManyBody,Eisert2018}
\begin{equation}
\rho_A=\frac{1}{Z}\int \mathrm{d}[\xi ]\mathrm{d}[\bar{\xi} ]\ee^{-\sum_{j}\bar{\xi}_j\xi_j} \bra{\{\xi_j\}}\rho_A\ket{\{\bar{\xi}_j\}} \ket{\{\xi_j\}}\bra{\{\bar{\xi}_j\}}.
\label{gra}
\end{equation} 
Here $\xi,\bar{\xi}$ are Grassman variables and $\ket{\xi}=\ee^{-\xi a^{\dagger}}\ket{0}$,$\ket{\bar{\xi}}=\bra{0}\ee^{- a^{\dagger}\bar{\xi}}$ are the related fermionic coherent states.
The partial time reversal, analog of Eq.~\eqref{eq:TRcoherentstate}, is \cite{shapourian2017partial}
\begin{equation}
    \ket{\xi}\bra{\bar{\xi}} \overset{\mathcal{T}}{\rightarrow}\ket{i \bar{\xi}}\bra{i \xi}.
    \label{TF}
\end{equation}
The partial time reversal $\rho_A^{R_1}$, obtained by acting with \eqref{TF} in \eqref{gra} only in $A_1$,  provides the fermionic negativity as ${\cal E}= \log{\rm Tr} |\rho_A^{R_1}|$, although its spectrum is not real in general \cite{shapourian2019twisted}. To have a more transparent interpretation of the fermionic negativity, an alternative partial transpose, called twisted fermionic partial transpose, has been defined as~\cite{shapourian2019twisted}
\begin{equation}
    \rho_A^{\tilde{R}_{1}}=\rho_A^{R_{1}}(-1)^{F_{A_1}},
\end{equation}
where $F_{A_1}=\sum_{j\in A_1}n_j$ is the number of fermions in the subsystem $A_1$. This new object has only real eigenvalues and the logarithmic negativity  
\begin{equation}\label{eq:lognegdef}
    \mathcal{E}=\log \mathrm{Tr}|\rho_A^{\tilde{R}_1}|.
\end{equation}
is a measure of the negativeness of the eigenvalues, exactly as for the bosonic partial transpose.
%
We define the negativity Hamiltonian related to $\rho_A^{R_{1}}$ as ${\cal N}_A$  and the one related to $\rho_A^{\tilde{R}_{1}}$ as $\tilde{\cal N}_A$.
\paragraph{Bisognano-Wichmann theorem -}
The BW theorem gives a general structure for the entanglement Hamiltonian of the ground state of a relativistic invariant quantum field theory with Hamiltonian density $H(\bf{x})$, when considering a bipartition between two half spaces of an infinite system. 
In formulas, considering a $d$-dimensional system, $\mathbf{x}=\{x_1,\dots,x_d\}$, and a partition $A=\{\mathbf{x}|x_1>0\}$, the EH of the ground state is
$
    H_A=2 \pi \int_{\mathbf{x}\in A} \mathrm{d} \mathbf{x} \; x_1 H(\mathbf{x})+c,
$
where $c$ is a normalization constant. 
This result does not depend on the dimensionality of the system or on any apriori knowledge of the ground state and can be applied to a large variety of systems and quantum phases. 
For conformal invariant theories, the BW theorem is easily generalized to some different geometries by conformal mappings  \cite{by-99,chm-11,wkzpv-13,ct-16}. 
This equivalence does not hold when $A$ is the union of two disjoint intervals, but, nevertheless, the EH for this geometry is known for $1+1$-dimensional free Dirac fermions \cite{ch-09}. 
In this case, it is possible to identify a local part in the entanglement Hamiltonian proportional to the energy density and a quasi-local part quadratic in the fermionic field. We will make explicit use of this example in the following. 
We will also check our analytical prediction against lattice simulations. %
In fact, the BW theorem can be used to construct approximate entanglement Hamiltonians for lattice models. This has been extensively investigated both for one- and two-dimensional models and it has been shown that the approximation provided by BW theorem allows to build entanglement Hamiltonians that encode all the relevant entanglement properties of the ground states~\cite{dalmonte2018quantum,Giudici2018,zcd-2020}.

\begin{figure}
     \centering
     \includegraphics[width=\linewidth]{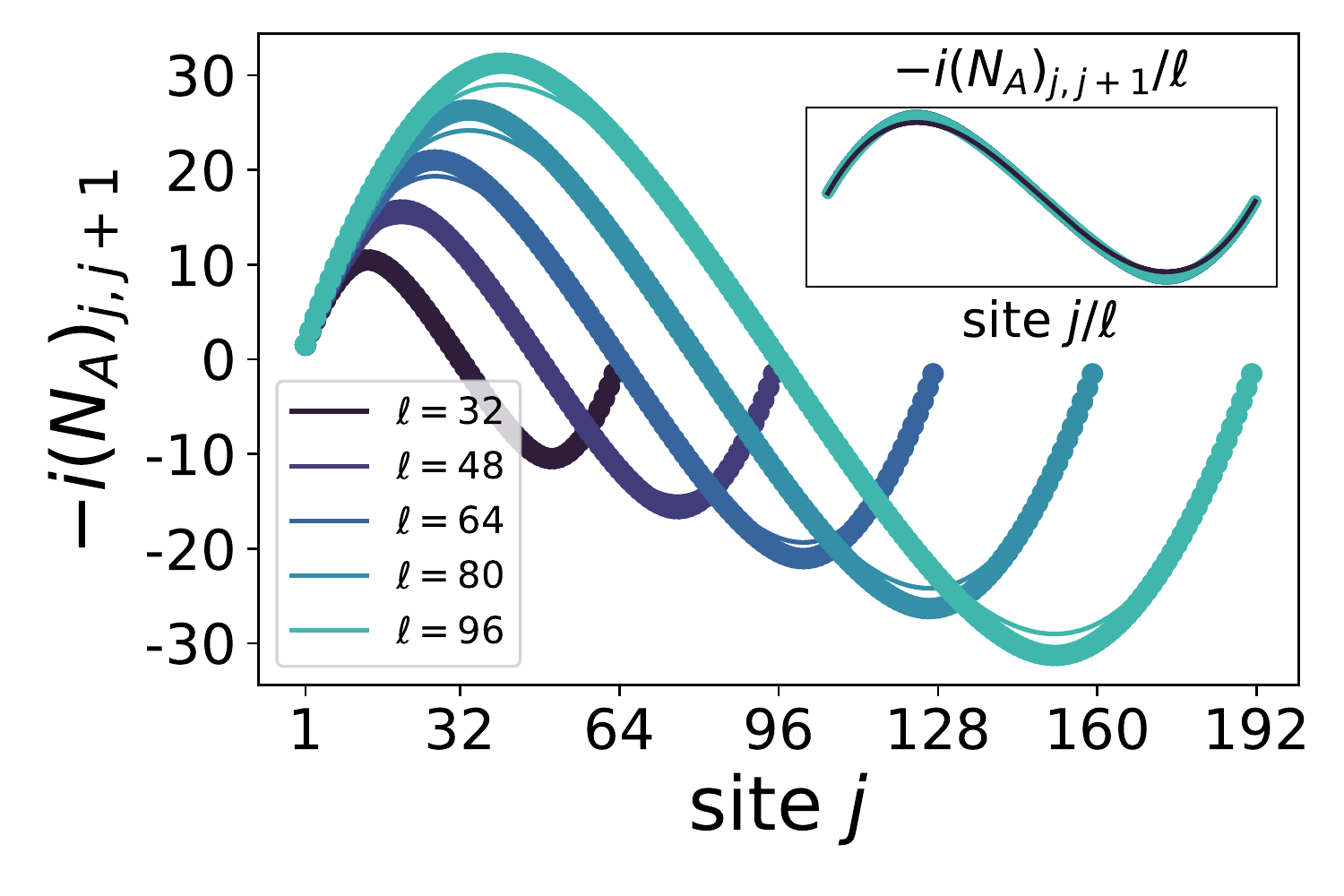}
     \includegraphics[width=\linewidth]{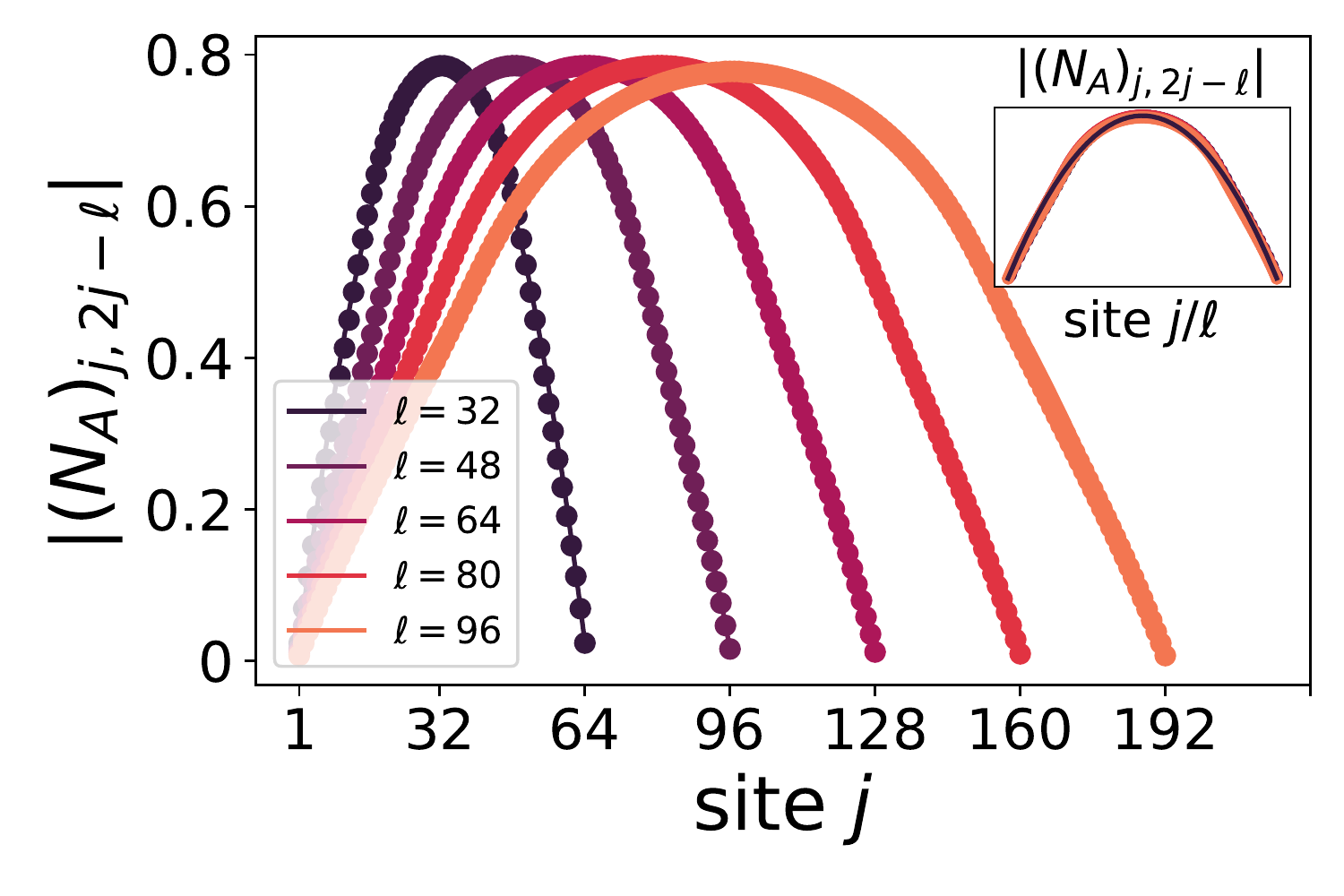}
     \caption{Benchmark of the analitycal prediction for the negativity Hamiltonian of a real fermion. We consider $A_1=[1,\ell], A_2=[\ell+1,2\ell]$ embedded in the infinite line. The symbols correspond to numerical data, while the solid lines to the discretized form of Eq. \eqref{eq:NH-f}. Upper panel: $N_{A,{\rm loc}}$. Lower panel: $|N_{A,{\rm q-loc}}|$. Inset: data collapse.}
     \label{fig:is_anti}
\end{figure}
\paragraph{ The Negativity Hamiltonian and its quasi-local structure -} 
To build the negativity Hamiltonian, we should first recall the path integral construction of the (bosonic) partial transpose  \cite{cct-12,cct-13}. 
The partial transposition  corresponds to the exchange of row and column indices in $A_1$ which naturally leads to a space inversion within $A_1$.
On a fundamental level, this fact can be deduced from CPT theorem. Indeed, the partial transposition is equivalent to a partial time reversal that, by CPT, is the same as a parity operation in the world-sheet combined with a charge conjugation. This second construction holds true also for $\rho_A^{R_1}$ in fermionic systems.


Therefore, starting from  the entanglement hamiltonian for two disjoint intervals  and doing a spatial inversion of the interval $A_1 = [a_1,b_1]$, one obtains the partial time reversal of the density matrix. 
Although this procedure is fully general, the entanglement hamiltonians of disjoint intervals are known only in few cases \cite{EH-f2,ch-09,CHP2018,mt-21,mt2-21,hollands,EH-disjoint2017,EH-disjoint2019}. In particular, starting from the EH for the massless real (Majorana) fermion $\Psi(x)$ \cite{ch-09}, 
$\Psi(x)=\begin{pmatrix}
\psi_1(x) \\
\psi_2(x)
\end{pmatrix}
$,
and performing this inversion, we get after simple algebra \cite{SM}
\begin{equation}
\begin{aligned}\label{eq:NH-f}
    &{\cal N}_A={\cal N}_{A,{\rm loc}}+i {\cal N}_{A,{\rm q-loc}}, \\
    &{\cal N}_{A,{\rm loc}}=2\pi \int_A \beta^{R}_{\rm loc}(x)T_{tt}(0,x) dx, \\ 
    &{\cal N}_{A,{\rm q-loc}}=2\pi \int_A \beta^R_{\rm q-loc}(x)T_{\rm q-loc}(x,{\bar{x}}^R(x)) dx,
\end{aligned}
\end{equation}
where 
\begin{equation}
\beta^R_{\rm loc}(x)=\frac{1}{w^{R} (x)'}, \qquad \beta^R_{\rm q-loc}(x)=\frac{\beta^R_{\rm loc}(\bar{x}^R(x))}{x-\bar{x}^R(x)},
\end{equation}
with
\begin{equation}\label{eq:xbar_wbar}
\begin{split}
    w^R(x) &=\log \left[-\frac{(x-b_1)(x-a_2)}{(x-a_1)(x-b_2)} \right],\\
    \bar{x}^R(x)&=\frac{(a_1b_2-b_1a_2)x+(a_1+b_2)b_1 a_2-(b_1+a_2)a_1b_2}{(a_1-b_1+b_2-a_2)x+b_1a_2-a_1b_2}.
\end{split}
\end{equation}
Here $T_{tt}(0,x)$ is the energy density  operator of the theory while $T_{\rm q-loc}(x,\bar{x})$ is a bilinear of the real fermionic fields, with  $x \in A_1$ and $\bar{x} \in A_2$ (and viceversa), i.e.
\begin{multline}\label{eq:Tnnlocal}
    T_{\rm q-loc}(x,y)\equiv i: (\psi_1(x)\psi_1(y) +\psi_2(x)\psi_2(y)
    ) :.
\end{multline}


The structure of Eq. \eqref{eq:NH-f} is very suggestive: it consists of a local term proportional to the energy density and an additional non local part given by a quadratic expression in the fermionic field. The latter, however, has a mild non-locality:  each point $x\in A_1$ is coupled to only a specific  $y={\bar x}^R\in A_2$ (that is a consequence of the mirror symmetry for equal intervals). Thus, following \cite{ch-09}, we refer to ${\cal N}_{A,{\rm q-loc}}$ as a quasi-local operator. Its existence is the reason of the imaginary components in the spectrum of ${\cal N}_A$, which is one characteristic treat of $\rho^{R_1}_A$.
The shape of $|{\cal N}_{A,{\rm q-loc}}|$ (see also Fig. \ref{fig:is_anti}) is compatible with the results of the negativity contour \cite{neg-ferm} suggesting that the largest contribution to the negativity comes from the boundary region between $A_1$ and $A_2.$ 
\begin{figure*}
\centering
     \includegraphics[width=0.325\linewidth]{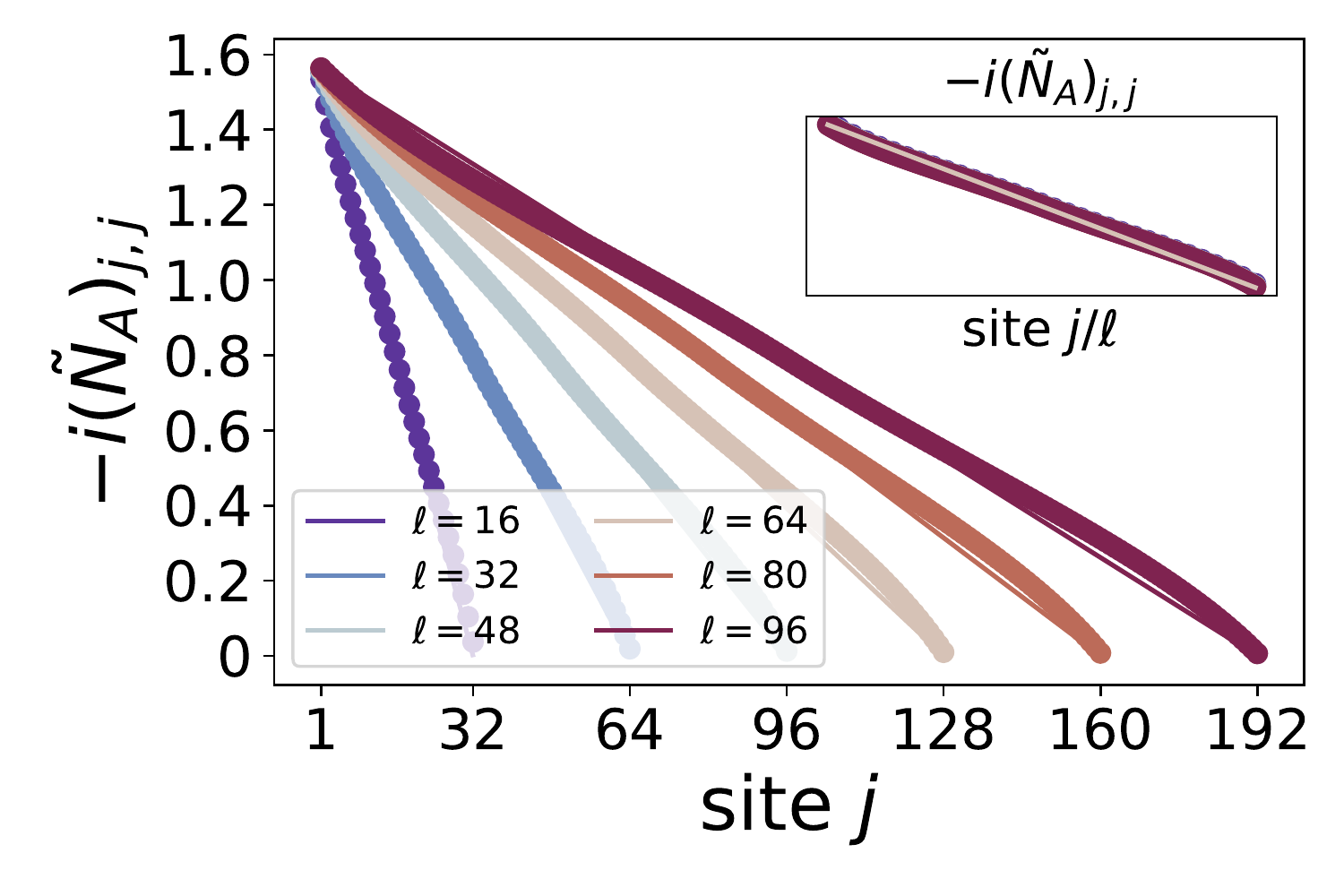}
     \includegraphics[width=0.325\linewidth]{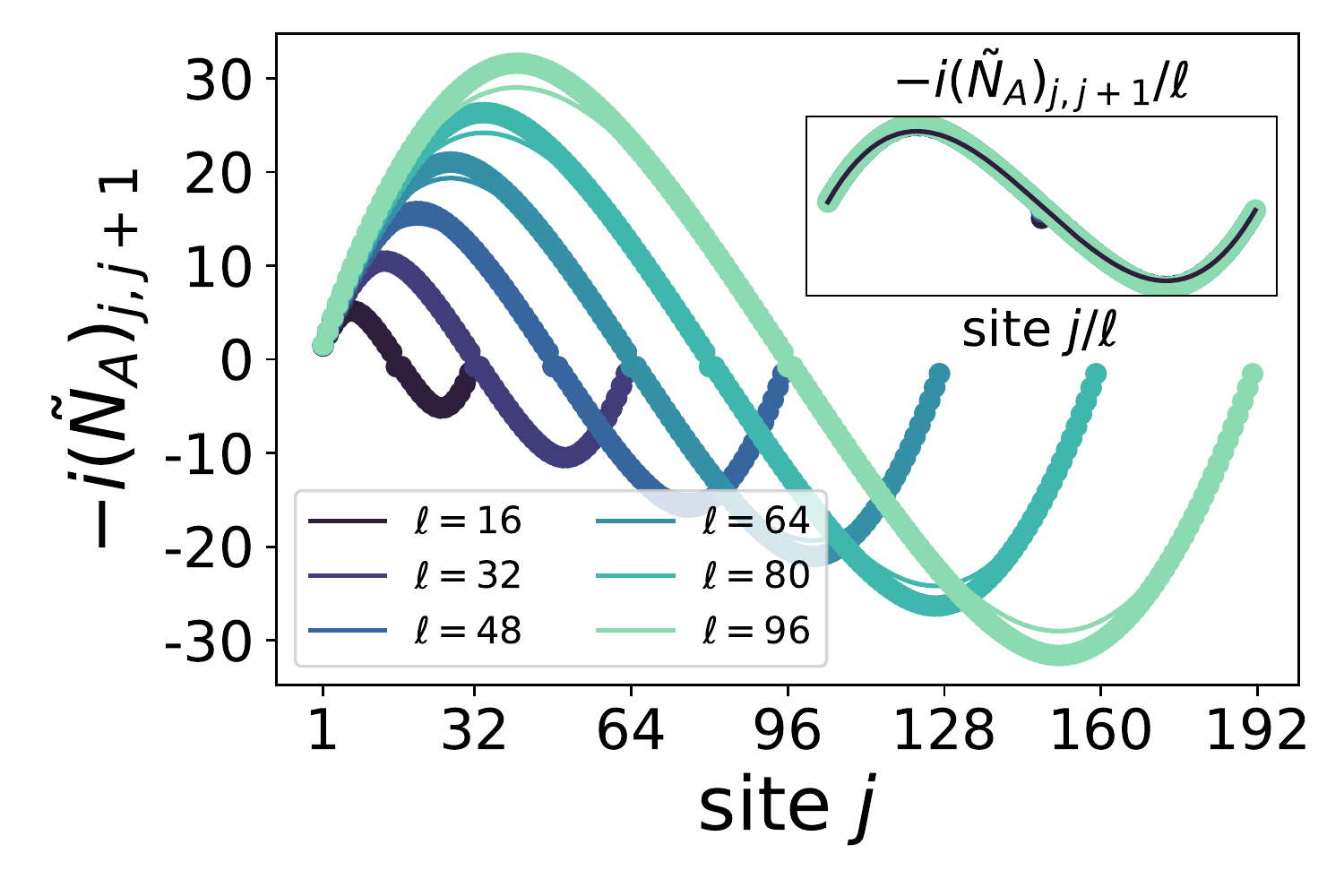}
     \includegraphics[width=0.325\linewidth]{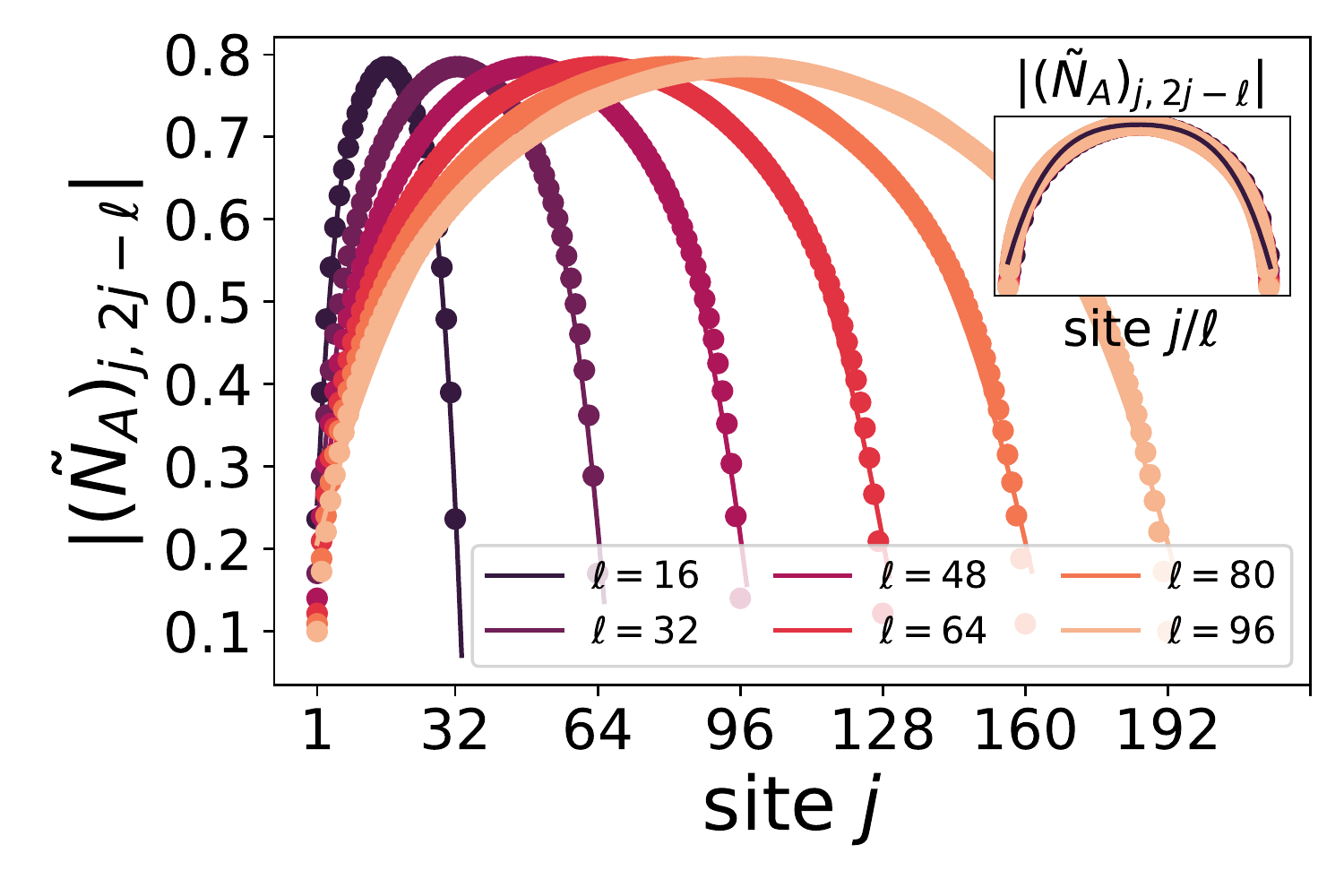}
     \caption{Benchmark of the analytic prediction for the Negativity Hamiltonian $\tilde{N}_A$. The symbols correspond to numerical data, while the solid lines correspond to the discretized form of Eq. \eqref{eq:NH-ftilde} for two adjacent intervals of length $\ell$. From left to right:  comparison $\tilde{N}_{A,{\rm loc}}$,  $\tilde{N}_{A,{\rm q-loc}}$ and $\tilde{N}_{A,{\rm q-loc}}$ with exact lattice simulations. Insets: data collapse. }
     \label{fig2:wtilde2}
\end{figure*}

To test the validity of Eq. \eqref{eq:NH-f}, we consider a lattice discretization of the Hamiltonian of free real fermions. 
Because of the gaussianity of $\rho_A^{R_1}$ \cite{shapourian2017partial},
the numerical evaluation of the negativity Hamiltonian  amounts to compute the single particle operator $N_A$ defined as ${\cal N_A}=\sum_{ij} (N_A)_{i,j} \psi_j \psi_i$,  related to the covariance matrix \cite{Peschel_2003,Peschel_2009}. We focus on two equal adjacent intervals $A=A_1 \cup A_2$ made up of $\ell$ sites labelled by $1 \leq j \leq 2\ell$. 
In this case, the point $\bar{x}^R$ in Eq. \eqref{eq:xbar_wbar} is just $\bar{x}^R=2\ell-x$ and so the quasi-local term lies entirely on the antidiagonal.
As a consequence, 
in Fig. \ref{fig:is_anti} we show only the subdiagonal $(N_{A})_{j,j+1}$ (a similar behaviour can be found for $(N_A)_{j+1,j}$) and the antidiagonal $(N_A)_{j,2\ell-j}$ which correspond, respectively, to the local and to the quasi-local parts of $N_A$.
The agreement between lattice exact and field-theoretical discretized $N_A$ is remarkable over the all parameter regime, and even for modest system sizes. Small discrepancies up to a few percent are present far from the boundaries: those have very little effects on the negativity, as they affect only very small (in absolute value)  eigenvalues of the partial transpose. 
We verified that the other  matrix elements of $N_A$ are negligible, in the sense that they are subleading as $\ell \to \infty$ (in the same sense as subleading terms in the EH are negligible, see Refs. ~\cite{dalmonte2018quantum,Giudici2018,zcd-2020,EH-l1,EH-l2,EH-l3,EH-1,EH-2}).

We have also studied the structure of negativity Hamiltonian $\tilde{\cal N}_A$ for two adjacent intervals of equal length, $\ell_1=\ell_2=\ell$. 
Although we did not manage to derive its form explicitly, we provide a conjecture that very accurately matches numerical data on the lattice. 
It reads
$ \tilde{\cal N}_A= \tilde{\cal N}_{A,{\rm diag}}+\tilde{\cal N}_{A,{\rm loc}}+ \tilde{\cal N}_{A,{\rm qloc}}$, with
\begin{equation}
\begin{split}\label{eq:NH-ftilde}
    &\tilde{\cal N}_{A,{\rm diag}}=2\pi i \int_A \tilde{\beta}_{\rm diag}(x) dx, \\ 
    &\tilde{\cal N}_{A,{\rm loc}}=2\pi  \int_A \tilde{\beta}_{\rm loc}(x) dx T_{tt}(0,x), \\ 
    &\tilde{\cal N}_{A,{\rm q-loc}}=2\pi \int_A \tilde{\beta}_{{\rm q-loc}}(x)T_{\rm q-loc}(x,{\bar{x}}^R) dx,
\end{split}
\end{equation}
where 
\begin{equation}
\begin{aligned}
&\tilde{\beta}_{\rm diag}(x)=\frac{1}{2}-\frac{x}{8\ell}, \\
&\tilde{\beta}_{\rm loc}=-\frac{ x (8 \ell^2 - 6 \ell x + x^2)}{8 \ell^2}, \\
&\tilde{\beta}_{\rm q-loc}(x)=4  \left(\frac{x - 2 \ell-\frac{1}{2} }{4 \ell}\right)^4 + 
\frac{1}{2}\left( \frac{x-2 \ell-\frac{1}2}{4 \ell}\right)^2 -\frac{1}{2},
\end{aligned}
\end{equation}
As a non-trivial test for the accuracy of this conjecture, 
we verified that it provides a logarithmic negativity that, as $\ell$ increases, approaches the exact numerical value (see Fig. \ref{fig:sketch1}).
We also benchmarked the analytical predictions from Eq. \eqref{eq:NH-ftilde} against exact computations, as shown in Fig. \ref{fig2:wtilde2}, for the one-particle NH, i.e. $\tilde{\cal N}_A=\sum_{ij} (\tilde{N}_A)_{i,j} \psi_j \psi_i$. Remarkably, the formulas above are in good agreement with simulations and, as already observed, the small discrepancies do not affect sizeably the logarithmic negativity approximation. The inset illustrates how results from different partition sizes collapse onto a single functional form, signaling scale invariance. 
A final comment concerns the spectrum of $\tilde{N}_A$: it consists of two parts $\{\lambda_j+i\pi\}, \lambda_j \in \mathbb{R}$ for $j=1, \cdots, 2\ell$ and $\{\lambda_j\}, \lambda_j \in \mathbb{R}$ for $j=2\ell+1, \cdots, 4\ell$.
By simple exponentiation, we get the eigenvalues of $\rho^{\tilde{R}_{1}}_A$, see SM. We can then trace back the appearance of negative eigenvalues in the spectrum of $\rho^{\tilde{R}_{1}}_A$
(and, as a consequence, of a non-zero negativity) to the presence of the factors $i \pi$ in $\tilde{N}_A$. 

Other tests of the analytic formulas for the negativity hamiltonian $N_A$, including different and disjoint intervals, are reported in the SM.

\paragraph{Discussion and outlook. -} In this work we initiated the study of the negativity Hamiltonian in many-body quantum systems. Although our field theoretical construction in terms of the EH of disjoint intervals is very general, its applicability relies crucially on the exact knowledge of the latter, that is not always available. 
We hope that this work will spark further studies on disjoint intervals's EH and, at the same time, the search for alternative constructions of $N_A$. 
We expect that the quasi-local structure of the negativity Hamiltonian can be generalised to other contexts, at least for free fermions, such as a single interval in an infinite system at finite temperature \cite{ct-16}, or two disjoint intervals in the presence of a point-like defect \cite{mt2-21}.
At present, it is unclear whether this quasi-local structure survives to finite interaction strengths and in higher dimensions.

Having established an explicit approximate functional form for the negativity Hamiltonian that is quasi-local opens up several possible applications. First, one could design experiments aimed at a direct realization of $N_A$: since the corresponding operators have simple functional form, this could be done by combining local tuning with tailor-engineered long-distance couplings similarly to what has already been proposed in the context of quantum chemistry simulations~\cite{Luengo2019}. Second, the local structure of $N_A$ paves the way for a direct reconstruction of partial transposes in experiments, utilizing, e.g., Hamiltonian reconstruction methods that have already been combined with the BW theorem~\cite{Kokail2021nature}. 
Both of these applications would allow a direct measurement of the negativity spectrum, something that is presently unachievable by any method other than full state tomography. Thirdly, it may be possible to design efficient classical or hybrid classical-quantum algorithms for the {\it ab initio} determination of $N_A$, similarly to what has been done for the EH following a BW inspired ansatz ~\cite{Parisen2018,Zhu2019,Kokail2021}. Having an explicit functional form could enable computations that are then not available otherwise - one example being quantum Monte Carlo algorithms aimed at computing the negativity utilizing meta-dynamics, similarly to what has been done in the context of the EH~\cite{TMS2020}.

We thank John Cardy for useful discussions. 
The work of M.D. and V.V. is partly supported by the ERC under grant number 758329 (AGEnTh), by the MIUR Programme FARE (MEPH), and by European Union's Horizon 2020 research and innovation programme under grant agreement No 817482 (Pasquans).
P.C. and S.M. acknowledge support from ERC under Consolidator grant number 771536 (NEMO).

S.M. and V.V. contributed equally to this manuscript.

\bibliography{biblio}

\beginsupplement

\section{From the Entanglement Hamiltonian and Bisognano Wichmann theorem to the Negativity Hamiltonian in field theory} \label{sec:EH_BW}

The calculation of the exact entanglement hamiltonian is in general a very difficult task. 
However, for conformal invariant field theories (CFTs) it is possible to generalize the Bisognano Wichmann (BW) result for a bipartition between two half spaces of an infinite system to different geometries \cite{by-99,chm-11,wkzpv-13,ct-16}. 

Let us consider the vacuum state of a $d$-dimensional Hamiltonian of a relativistic quantum field theory,
\begin{equation}
    H=\int_{\mathbb{R}^d} d^dx \, h(x) 
\end{equation}
and a subsystem $A$ which consists of the degrees of freedom in a half-space, $x_1>0$. The BW theorem guarantees that the entanglement Hamiltonian $H_A$ can be expressed as an integral of the Hamiltonian density $h(x)$
\begin{equation}\label{eq:Ha-half}
    H_A=\frac{2\pi}{v}\int_{A}d^{d-1}x\, x_1\, h(x),
\end{equation}
where, from now on, we fix the velocity $v=1$.
There are other examples in the ground-state of a 1+1 dimensional CFT in which $H_A$ can be written as a local integral over the Hamiltonian density. They include the case of a single interval $A=(0,\ell)$ in an infinite system, and its generalizations to finite size or finite temperature \cite{ct-16}. In these cases, $H_A$ takes the form
\begin{equation}\label{eq:Ha-map}
    H_A=2\pi \int_A dx \frac{h(x)}{f'(x)},
\end{equation}
where $f'(x)$ is the conformal mapping from the Euclidean space-time to a rectangle with height $2\pi$ and width $2\log(\ell/\epsilon)$, $\epsilon$ UV cutoff. More concretely, some mappings are 
\begin{itemize}
    \item finite interval in an infinite system: $$f(x)=\log \frac{x}{\ell-x};$$
     \item finite interval in a finite system: $$f(x)=\log \frac{e^{2 \pi i x/L} - 1}{e^{2 \pi i \ell/L}-e^{2 \pi i x/L}}.$$
\end{itemize}
Despite the result of Bisognano and Wichmann and the conformal symmetry allow to compute the aforementioned modular Hamiltonians, in general it is not an easy task to get analytic expressions, even in CFTs. One of these examples is the modular Hamiltonian for the ground state of the free $1+1$ dimensional massless Dirac fermion for several disjoint intervals on the infinite line \cite{CHP2018,ch-09,mt-21}. We have already discussed its peculiar structure in the main text and here we report the explicit analytical expression. The massless $1+1$ dimensional Dirac field $\psi(t, x)$ is a doublet made by the two complex fields
\begin{equation}\label{eq:diracspinor}
\psi(t,x)=\begin{pmatrix}
\psi_1(t,x) \\
\psi_2(t,x)
\end{pmatrix}.
\end{equation}
The normal ordered component of the energy-momentum tensor of the Dirac field corresponding to the energy density reads
\begin{equation}\label{eq:energydensity}
   T_{tt}(t,x)\equiv \frac{i}{2}:[\left((\partial_x\psi_1^*)\psi_1-\psi_1^*\partial_x\psi_1\right)(x+t)\\- \left((\partial_x\psi_2^*)\psi_2-\psi_2^*\partial_x\psi_2\right)(x-t)]:.
\end{equation}
The modular Hamiltonian for two disjoint intervals $A \equiv [a_1, b_1] \cup [a_2, b_2]$ on the line can be written as the sum $ H_A = H_{\rm loc}+H_{{\rm q-loc}}$, where the local term $H_{\rm loc}$ and the quasi-local term $H_{{\rm q-loc}}$ are defined respectively as
\begin{equation}\label{eq:EH-disjoint}
\begin{aligned}
    &H_{\rm loc}=2\pi \int_A \beta_{\rm loc}(x)T_{tt}(0,x) dx, \\ 
    &H_{{\rm q-loc}}=2\pi \int_A \beta_{{\rm q-loc}}(x)T_{{\rm q-loc}}(0,x,\bar{x}) dx,
\end{aligned}
\end{equation}
with $T_{tt}(0,x)$ the energy density in Eq. \eqref{eq:energydensity}, while $T_{{\rm q-loc}}(0,x,y)$ is given by
\begin{equation}\label{eq:Tnnlocal}
    T_{{\rm q-loc}}(t,x,y)\equiv \frac{i}{2}:[ (\psi_1^*(x+t)\psi_1(y+t)
    -\psi_1^*(y+t)\psi_1(x+t))+(\psi_2^*(x-t)\psi_2(y-t)
    -\psi_2^*(y-t)\psi_2(x-t)) ]:.
\end{equation}
Here the asterisk denotes the Hermitian conjugation. The other functions in Eq. \eqref{eq:EH-disjoint} can be written as
\begin{equation}
    \beta_{\rm loc}(x)=\frac{1}{w'(x)} \qquad \beta_{{\rm q-loc}}(x)=\frac{\beta_{\rm loc}(\bar{x}(x))}{x-\bar{x}(x)},
\end{equation}
with
\begin{equation}\label{eq:xbar}
\begin{aligned}
    w(x) &=\log \left[-\frac{(x-a_1)(x-a_2)}{(x-b_1)(x-b_2)} \right],\\
    \bar{x}(x)&=\frac{(b_1b_2-a_1a_2)x+(b_1+b_2)a_1 a_2-(a_1+a_2)b_1b_2}{(b_1-a_1+b_2-a_2)x+a_1a_2-b_1b_2}.
\end{aligned}
\end{equation}
Here $x$ and $\bar{x}(x)$ belong to different intervals in $A$ (if $x\in A_1$ then $\bar x\in A_2$ and viceversa). 
In the limit $b_1 \to a_2$ we get back a single interval and so the quasi-local part vanishes and we recover the result in Eq. \eqref{eq:Ha-map}, where now $A=[a_1,b_2]$. 

\begin{figure}
\centering  
    \includegraphics[width=0.58\linewidth]{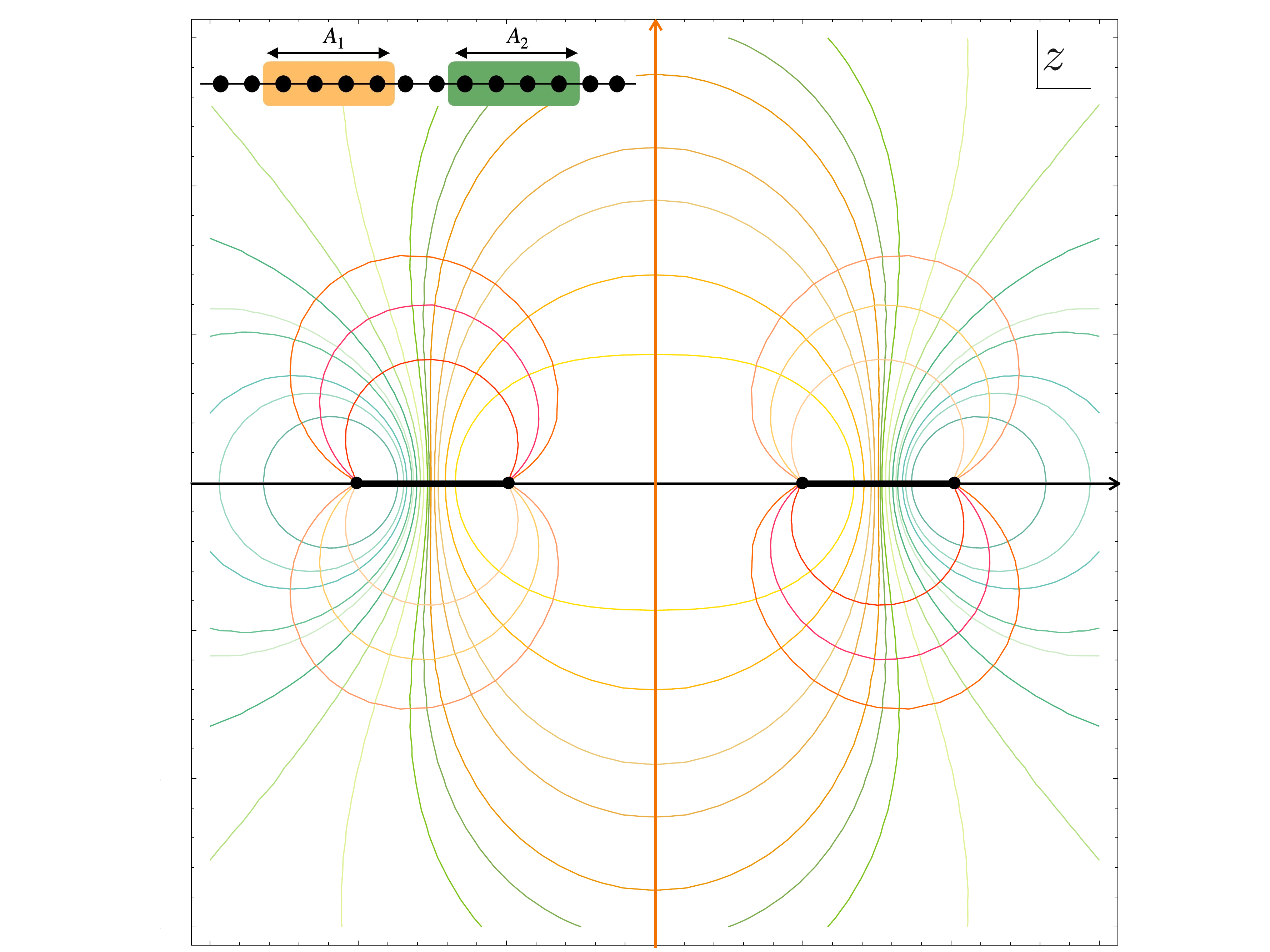}
    \caption{Contour plot of the conformal mapping $w^R(z)$ defined in Eq. (11) of the main text, where $z=x+iy$.  Here $A$ consists of two disjoint intervals whose endpoints are the black dots on the horizontal axis. The absence of singular points along the imaginary axis is the main difference with respect to the conformal mapping $w(z)$ in Eq. \eqref{eq:xbar}.}
	\label{fig:mapping}
\end{figure}

The entanglement Hamiltonian in Eq. \eqref{eq:EH-disjoint} is the starting point to obtain an analytical expression for the negativity Hamiltonian $\mathcal{N_A}$. 
As explained in the main text, in the path integral representation the partial transposition has the net effect to perform a spatial inversion within $A_1$ plus a charge conjugation. This implies that the negativity Hamiltonian $\mathcal{N}_A$ can be obtained from the entanglement Hamiltonian $H_A$ by inverting the endpoints $a_1 \leftrightarrow b_1$ in the expression for $H_A$, Eq. \eqref{eq:EH-disjoint}. 
Furthermore, since under partial time reversal $\psi(x) \to i \psi(x)$ if $x \in A_1$, the term  $T_{{\rm q-loc}}(0,x,\bar{x})$ defined in Eq.~\eqref{eq:Tnnlocal} gets an $i$ prefactor, because if $x \in A_1$ then $\bar{x} \in A_2$ (and viceversa). 
To sum up, we get the following expression for the negativity Hamiltonian of a Dirac field
\begin{equation}
\begin{aligned}\label{eq:NH-fcomplex}
    &{\cal N}_A={\cal N}_{A,{\rm loc}}+i {\cal N}_{A,{\rm q-loc}}, \\
    &{\cal N}_{A,{\rm loc}}=2\pi \int_A \beta^{R}_{\rm loc}(x)T_{tt}(0,x) dx, \\
    &{\cal N}_{A,{\rm q-loc}}=2\pi \int_A \beta^R_{\rm q-loc}(x)T_{\rm q-loc}(0,x,{\bar{x}}^R(x)) dx,
\end{aligned}
\end{equation}
where $\beta^R_{\rm loc},\beta^R_{\rm q-loc},{\bar{x}}^R(x)$ are given in Eq. (10)-(11)  of the main text  and they are obtained by switching $a_1$ and $b_1$ in the definitions of $\beta_{\rm loc},\beta_{\rm q-loc},{\bar{x}}(x)$ given above. In Fig. \ref{fig:mapping}, we show a contour plot of the real and imaginary part of $w^R(x)$. The curves $\mathrm{Im}(w^R(x))$ do not exhibit singular points for which the mapping fails to
be conformal, contrarily to what happens for $\mathrm{Im}(w(x))$ and showed in Fig. 9 of \cite{ct-16}.
The presence of this singularity prevents from applying the BW theorem with a conformal mapping given by $w(x)$ \cite{ct-16}. 
The operators $T_{tt}, T_{\rm q-loc}$ are the ones in Eq. \eqref{eq:energydensity}, \eqref{eq:Tnnlocal}, respectively.

At this point, the result for real (Majorana) fermions can be obtained without any further effort. Indeed, the Dirac spinor in Eq. \eqref{eq:diracspinor} can be written in terms of two Majorana spinors (real fermions). Rewriting the  negativity (entanglement) hamiltonian in terms of these components, the mixed term cancel and ${\cal N}_A$ ($H_A$) is the sum of the two negativity (entanglement) hamiltonians for each real component of the complex field. As a consequence Eq. \eqref{eq:EH-disjoint} is valid also for real massless fermions in 1+1 dimensions with  $T_{tt}(0,x)$ the energy density of the real fermions and $ T_{{\rm q-loc}}(0,x,y)$ given by Eq. (12) of the main text.
There, we have explicitly reported only the expression of $\mathcal{N}_A$ and $\mathcal{\tilde{N}}_A$ for Majorana, while in Eq. \eqref{eq:NH-fcomplex} we report the explicit expression of $\mathcal{N}_A$ for Dirac. The functional form of $\beta^R_{\rm loc},\beta^R_{\rm q-loc},{\bar{x}}^R(x)$ is the same and only the definition of $T_{tt}(0,x)$ and $ T_{{\rm q-loc}}(t,x,y)$ in terms of a real or complex fermionic field change. 

\section{Partial transpose of bosonic and fermionic systems}\label{sec:partial_transpose}

To work with partially transposed density matrices, we briefly introduced in the main text the partial transpose transformation for both bosonic and fermionic systems. Here we will review some results on the subject, and will discuss the case of Gaussian states for which the correlation matrices suffice to compute all the properties of the systems~\cite{ez-15,shapourian2017partial,shapourian2019twisted}. The techniques we present here are the ones we used to benchmark our predictions for $\mathcal{N_A}$ and $\mathcal{\tilde{N}_A}$.

For a bosonic system, the partial transpose of the reduced density matrix $\rho^{T_{A_1}}$ with respect to $A_1$ is defined by performing a standard transposition in $\mathcal{H}_{A_1}$, i.e. exchanging the matrix elements in $A_1$,
\begin{equation}\label{eq:bosonic_PT}
\rho^{T_{1}}_A=(T_{A_1} \otimes \mathds{1}_{A_2})\rho=\\
=\sum_{i,j,k,l} \bra{e^{A_1}_k,e^{A_2}_j}\rho_A \ket{e^{A_1}_i,e^{A_2}_l} \ket{e^{A_1}_i,e^{A_2}_j} \bra{e^{A_1}_k,e^{A_2}_l}.
\end{equation}
This definition can also be expressed in terms of the Wigner distribution functions $W(q,p)$ \cite{Simon2000}. 
The same is not true for fermions because of the anticommutation relations.

To construct the partial transpose for fermionic systems,
let us start by writing the density matrix in terms of Majorana operators $c_j$, which are defined in terms of the fermionic operators $a_j$ obeying $\{a^{\dagger}_k,a_j\}=\delta_{kj}$ as
\begin{equation}
\begin{cases}\label{eq:mojorana}
    c_{2j-1}=a_j+a^{\dagger}_j,\\   c_{2j}=i(a_j-a^{\dagger}_j).
\end{cases}
\end{equation}
We consider a system $\mathcal{S}=A_1\cup A_2$ and denote with the subscripts $\{m_1,\dots,m_{l_1}\}$ the operators in the subset $A_1$ and with $\{n_1,\dots,n_{l_2}\}$ the ones in the subset $A_2$; here $l_1$($l_2$) corresponds to the number of sites in subsystem $A_1$($A_2$). One can write~\cite{ez-15}:
\begin{equation}
    \rho_A=\sum_{\bm{\kappa},\bm{\tau}}w_{\bm{\kappa},\bm{\tau}}c_{m_1}^{\kappa_1}\dots c_{m_{2l_1}}^{\kappa_{2l_1}}c_{n_1}^{\tau_1}\dots c_{n_{2l_2}}^{\tau_{2l_2}}
\end{equation}
where we defined $\bm{\kappa}=(\kappa_1,\dots,\kappa_{2l_1})$ and $\bm{\tau}=(\tau_1,\dots,\tau_{2l_2})$ with $\kappa_j, \tau_j=0,1$. 
We define the moduli  $|\kappa|=\sum_{j=1}^{2l_1} \kappa_j$ and $|\bm\tau|=\sum_{j=1}^{2l_2}\tau_j$.
Since the physical fermionic states must commute with the parity operator one has that the sum of the moduli of $\bm{\kappa}$ and $\bm{\tau}$ must be even.
The partial transpose~\eqref{eq:bosonic_PT} leaves unaltered the state in $A_2$ and exchanges the states in $A_1$ as
\begin{equation}
    \rho^{T_{1}}_A=\sum_{\bm{\kappa},\bm{\tau}}(-1)^{f(\bm{\kappa})}w_{\bm{\kappa},\bm{\tau}}c_{m_1}^{\kappa_1}\dots c_{m_{2l_1}}^{\kappa_{2l_1}}c_{n_1}^{\tau_1}\dots c_{n_{2l_2}}^{\tau_{2l_2}}
    \label{TT1}
\end{equation}
where
\begin{equation}
    (-1)^{f(\bm{\kappa})}=\begin{cases}
    0 \;\;\;\; |\bm{\kappa}| \; \textrm{mod}\; 4 \in \{0,3\}\\
    1 \;\;\;\; |\bm{\kappa}| \; \textrm{mod}\;  4 \in \{1,2\}\\
    \end{cases}.
\end{equation}
The easiest way to see this is to perform the partial transpose in the occupation number basis and then write the density matrix in terms of Majorana operators.

Let us now consider a Gaussian state that can be written in the form
\begin{equation}
    \rho_A=\frac{1}{Z}\ee^{\frac{1}{4}\sum_{kl}W_{kl}c_kc_l},
\end{equation}
where $c_k$ are fermionic Majorana operators and $W$ is a $2\ell \times 2\ell$ matrix ($\ell$ size of the system described by $\rho$), with eigenvalues $\in \mathbb{R}$. The latter is related to the correlation matrix $\Gamma$ (i.e. the matrix with elements $\Gamma_{i,j}=\frac{1}{2}\langle [c_i,c_j]\rangle$) by the relation
\begin{equation}\label{eq:GammaWrelation}
\Gamma=\tanh{\frac{W}{2}}.
\end{equation}
Here $\Gamma$ has eigenvalues between $[-1,1]$.
It is convenient to introduce the block structure of $\Gamma$ as
\begin{equation}
    \Gamma=\begin{pmatrix}
    \Gamma_{A_1A_1} &\Gamma_{A_1A_2}\\
    \Gamma_{A_2A_1} & \Gamma_{A_2A_2}
    \end{pmatrix}.
\label{Gammablock}    
\end{equation}
Using Eq. \eqref{TT1} it can be shown that \cite{ez-15}
\begin{equation}\label{eq:standard_partial_transpose}
    \rho_A^{T_{1}}=\frac{1-i}{2}O_{+}+\frac{1+i}{2}O_{-}
\end{equation}
where $O_{\pm}=\sum_{\bm{\kappa},\bm{\tau}}o^{\pm}_{\bm{\kappa},\bm{\tau}}c_{m_1}^{\kappa_1}\dots c_{m_{2l_1}}^{\kappa_{2l_1}}c_{n_1}^{\tau_1}\dots c_{n_{2l_2}}^{\tau_{2l_2}}$ with 
\begin{equation}
    o^{\pm}_{\bm{\kappa},\bm{\tau}}=
\begin{cases}
\pm i (-1)^{\frac{|\bm{\kappa}|-1}{2}}w_{\bm{\kappa},\bm{\tau}}\;\;\; |\bm{\kappa}|\; \textrm{odd}\\
i (-1)^{\frac{|\bm{\kappa}|}{2}}w_{\bm{\kappa},\bm{\tau}}\;\;\qquad |\bm{\kappa}|\; \textrm{even}.
\end{cases}
\end{equation}
The operators $O_{\pm}$ are  Gaussian and can be written as
\begin{equation}
    O_{+}=O_{-}^{\dagger}=\frac{1}{Z}\ee^{\frac{1}{4}\sum_{kl}(N_{A})_{kl}c_kc_l }
\end{equation}
where $N_A$ is related (as in Eq.~\eqref{eq:GammaWrelation}) to the correlation matrix  $\Gamma_{+}$ defined according to the following equation:
\begin{equation}\label{eq:gammap}
    \Gamma_{+}=\begin{pmatrix} -\Gamma_{A_1A_1} &  i \Gamma_{A_1A_2} \\  i \Gamma_{A_2A_1} & \Gamma_{A_2A_2} \end{pmatrix}.
\end{equation}
It is clear that the partially transposed reduced density matrix \eqref{eq:standard_partial_transpose} is not a Gaussian operator, but rather the sum of two of them.
Even more troubling, it does not satisfy additivity nor subadditivity and fails to capture, for this reason, some topological features of fermionic Majorana systems such as the entanglement due to zero-energy modes in  Kitaev's chain \cite{shapourian2017partial}.

For all the above reasons, a different partial transpose has been introduced for fermionic systems starting from the analogy with the time-reversal transformation~\cite{shapourian2017partial,shapourian2019twisted,ShapourianManyBody}: we have already understood that the action of the fermionic partial transpose ($R$) is not just exchanging bra and ket but also multiply them by $i$ (see Eq. 6 of the main text). 
This definition can be readily generalized to multi-particle states.
Considering a system $\mathcal{S}=A_1 \cup A_2$ one has
\begin{equation}
    (\ket{\{\xi_j\}_{j\in{A_1}}\{\xi_j\}_{j\in{A_2}}}\bra{\{\bar{\chi}_j\}_{j\in{A_1}}\{\bar{\chi}_j\}_{j\in{A_2}}})^{R_{1}} =\ket{\{i \bar{\chi}_j\}_{j\in{A_1}}\{\xi_j\}_{j\in{A_2}}}\bra{\{i \xi_j\}_{j\in{A_1}}\{\bar{\chi}_j\}_{j\in{A_2}}},
\end{equation}
with obvious meaning of all the actors in the formula. In the occupation number basis, the above equation reads~\cite{shapourian2017partial}
\begin{equation}\label{eq:fermionictimereversal}
   (\ket{\{n_j\}_{j\in{A_1}}\{n_j\}_{j\in{A_2}}}\bra{\{\bar{n}_j\}_{j\in{A_1}}\{\bar{n}_j\}_{j\in{A_2}}})^{R_{1}} 
    =(-1)^{\phi(\{n_j,\bar{n}_j\})}\times \ket{\{\bar{n}_j\}_{j\in{A_1}}\{n_j\}_{j\in{A_2}}}\bra{\{n_j\}_{j\in{A_1}}\{\bar{n}_j\}_{j\in{A_2}}}.
\end{equation}
Here the term $\phi(\{n_j,\bar{n}_j\})$ is a phase factor depending on the occupation number 
\begin{equation}\label{eq:phasefactor}
    \phi(\{n_j,\bar{n}_j\})=\frac{\tau_{A_1}(\tau_{A_1}+2)}{2}+\frac{\bar{\tau}_{A_1}(\bar{\tau}_{A_1}+2)}{2}+\tau_{A_2}\bar{\tau}_{A_2}+\\
    +\tau_{A_1}\tau_{A_2}+\bar{\tau}_{A_1}\bar{\tau}_{A_2}+(\bar{\tau}_{A_1}+\bar{\tau}_{A_2})(\tau_{A_1}+\tau_{A_2}),
\end{equation}
with $\tau_{A_1}=\sum_{j\in A_1}n_j$ ($\tau_{A_2}=\sum_{j\in A_2}n_j$) and $\bar{\tau}_{A_1}=\sum_{j\in A_1}\bar{n}_j$ ($\bar{\tau}_{A_2}=\sum_{j\in A_2}\bar{n}_j$). Hence the definition in Eq.~\eqref{eq:fermionictimereversal} is equivalent to a standard partial transposition up to phase factor depending on the parity of the two subsystems, as in Eq. \eqref{eq:phasefactor}. In terms of Majorana operators,
the transformation in Eq.~\eqref{eq:fermionictimereversal} can be rewritten as
\begin{equation}\label{eq:fermionic_partial_transpose}
    \rho^{R_{1}}_A=\sum_{|\bm{\kappa}|+|\bm{\tau}| even} w_{\bm{\kappa},\bm{\tau}}i^{|\bm{\kappa}|}c_{m_1}^{\kappa_{m_1}}\cdots c_{m_{2l_1}}^{\kappa_{m_{2l_1}}}c_{n_1}^{\tau_{n_1}}\cdots c_{n_{2l_2}}^{\tau_{n_{2l_2}}}
\end{equation}
where we used the notation $c^0_x=\mathds{1}$, $c^1_x=c_x$. 
The matrix $\rho^{R_{1}}_A$ satisfies three necessary properties for a partial transposition:
\begin{enumerate}
    \item $(\rho_A^{R_{1}})^{R_{2}}=\rho_A^R$,
    \item $(\rho^{R_{1}}_A)^{R_{1}}=\rho_A$,
    \item $(\rho_{1}\otimes \rho_{2}\cdots \rho_{n})^{R_{1}}=\left(\rho_{1}^{R_{1}}\otimes \rho_{2}^{R_{1}}\otimes \cdots \otimes\rho_{n}^{R_{1}}\right)$.
\end{enumerate}
Notice that $\rho_A^{R_{A_1}}$ is nothing but $O_{+}$ in Eq.~\eqref{eq:standard_partial_transpose}.
This density operator is not Hermitian and, in general, has complex eigenvalues.
Nevertheless, one can still define a  fermionic logarithmic negativity as \cite{shapourian2017partial}
\begin{equation}
    \mathcal{E}=\log \mathrm{Tr}\sqrt{(\rho_A^{R_1})^{\dagger}\rho_A^{R_1}}
\label{Fneg1}
\end{equation}
where the object $(\rho_A^{R_1})^{\dagger}\rho_A^{R_1}$ is Hermitian and its spectrum is positive. 
In spite of the name, the fermionic negativity has nothing to do with the negativeness of the spectrum of $(\rho_A^{R_1})$. It can be however proved that is a proper entanglement monotone \cite{ryu-monotone19} and it has been shown that can detect entanglement when the standard negativity fails \cite{shapourian2017partial}. 
It has been shown in \cite{shapourian2019twisted} that, in general, there is a freedom in the definition of the partial transpose operation. This leads to an alternative definition for the fermionic negativity \eqref{Fneg1} given by
\begin{equation}
    \mathcal{E}=\log \mathrm{Tr}|\rho_A^{\tilde R_1}|.
\label{Efe2}
\end{equation}
Given that the spectrum of $\rho_A^{\tilde R_1}$ is real, the fermionic negativity is a measure of the negativeness of the eigenvalues of the latter.






In the case of Gaussian states, also the relation between $\rho^{\tilde{R}_{1}}_A$ and $\rho^{R_{1}}_A$ simplifies and one can write
\begin{equation}
    \ee^{\tilde{N}_{A}}=\frac{\mathds{1}+\Gamma_{+}}{\mathds{1}-\Gamma_{+}}U_{A_1},
\end{equation}
where the matrix $U_A=-\mathds{1}_{A_1} \oplus \mathds{1}_{A_2}$ is related to the transformation $(-1)^{F_{A_1}}$. 

A last comment concerns the structure of the spectrum of $\rho^{\tilde{R}_{1}}_A$.
The eigenvalues of $\tilde \Gamma$ are of the form 
$\pm \tilde \nu_j$. 
In terms of these, the density matrix $\rho^{\tilde{R}_{A_1}}$ can be brought into a diagonal form 
\begin{equation}\label{eq:prod}
    \rho^{\tilde{R}_{1}}_A=\frac{1}{Z}\prod_{j=1}^{\ell_1+\ell_2}\frac{1+i \tilde{\nu}_jd_{2j}d_{2j-1}}{2},
    \end{equation}
where $d_j$ are a set of real fermionic operators. 
Since the eigenvalues of $d_{2j}d_{2j-1}$ are $\pm i$, the $2^\ell$ eigenvalues of $\rho^{\tilde{R}_{A_1}}$ are given by all possible products of $\frac{1\pm \tilde \nu_j}{2}$.
Hence, in order to have negative eigenvalues a necessary conditions is that some $\nu_j$ are larger than 1. 
Focusing now, for practical reasons, to the case of two intervals of length $\ell_1$ and $\ell_2$ respectively, we anticipated in the main text that the eigenvalues of $\tilde{N}_{A}$ are either real or real $+i\pi$. To be more precise, the spectrum of $\tilde{N}_{A}$ is of the form $\{\lambda_j+i\pi\}, \lambda_j \in \mathbb{R}$ for $j=1, \cdots, 2\ell_1$ and $\{\lambda_j\}, \lambda_j \in \mathbb{R}$ for $j=2\ell_1+1, \cdots, 2(\ell_1+\ell_2)$. The presence of the terms $i \pi$ is what determines the negative eigenvalues in the spectrum of $\rho^{\tilde{R}_{1}}_A$. 
Indeed, using the relation
\begin{equation}
    \tilde{\Gamma}=\tanh \frac{\tilde{N}_A}{2},
\end{equation}
if the eigenvalues are $\lambda_j+i\pi$, then  $\nu_j=\tanh(\lambda_j/2+i\pi/2)=\coth(\lambda_j/2)$, and so $\nu_j=\coth(\lambda_j/2)>1$.

We finally mention that 
the fermionic logarithmic negativity in Eq. \eqref{Efe2} can be computed as
\begin{equation}\label{eq:fneg}
   \mathcal{E}=\sum_{j=1}^{\ell_1+\ell_2}\log\left[\Big| \frac{1-\tilde{\nu}_j}{2}\Big|+ \Big| \frac{1+\tilde{\nu}_j}{2}\Big|\right] +\log \mathrm{Tr}(\rho_A^{\tilde{R}_{1}}),
\end{equation}
where $\mathrm{Tr}(\rho_A^{\tilde{R}_{1}})=\sqrt{\mathrm{det}\Gamma_{A_1A_1}}$. Note that the sum is over half of the eigenvalues of $\tilde{\Gamma}$.

\section{Lattice Negativity Hamiltonian and numerical checks}\label{sec:negativity_hamiltonian}

\begin{figure}
\centering  
    \includegraphics[width=0.48\linewidth]{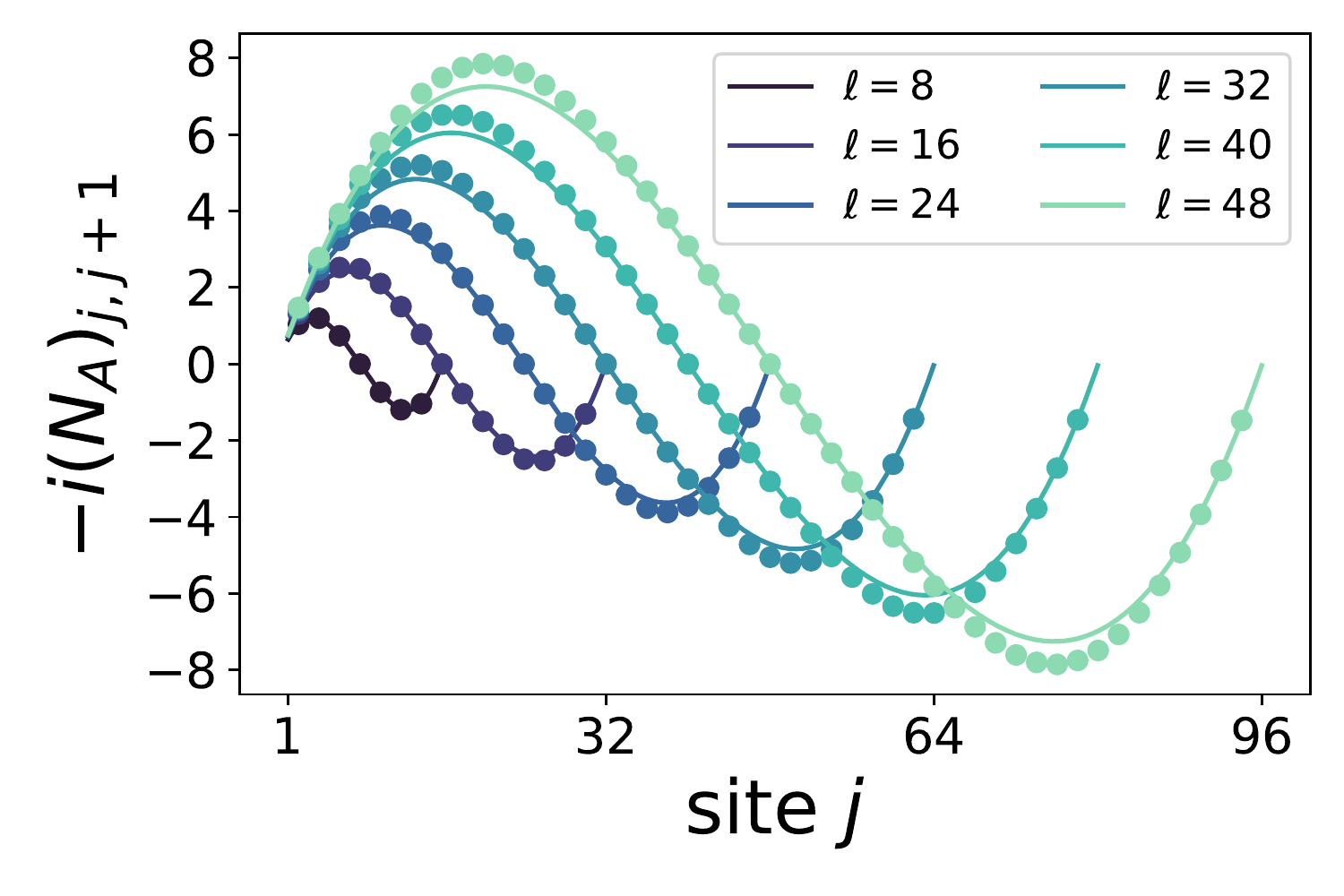}
    \includegraphics[width=0.48\linewidth]{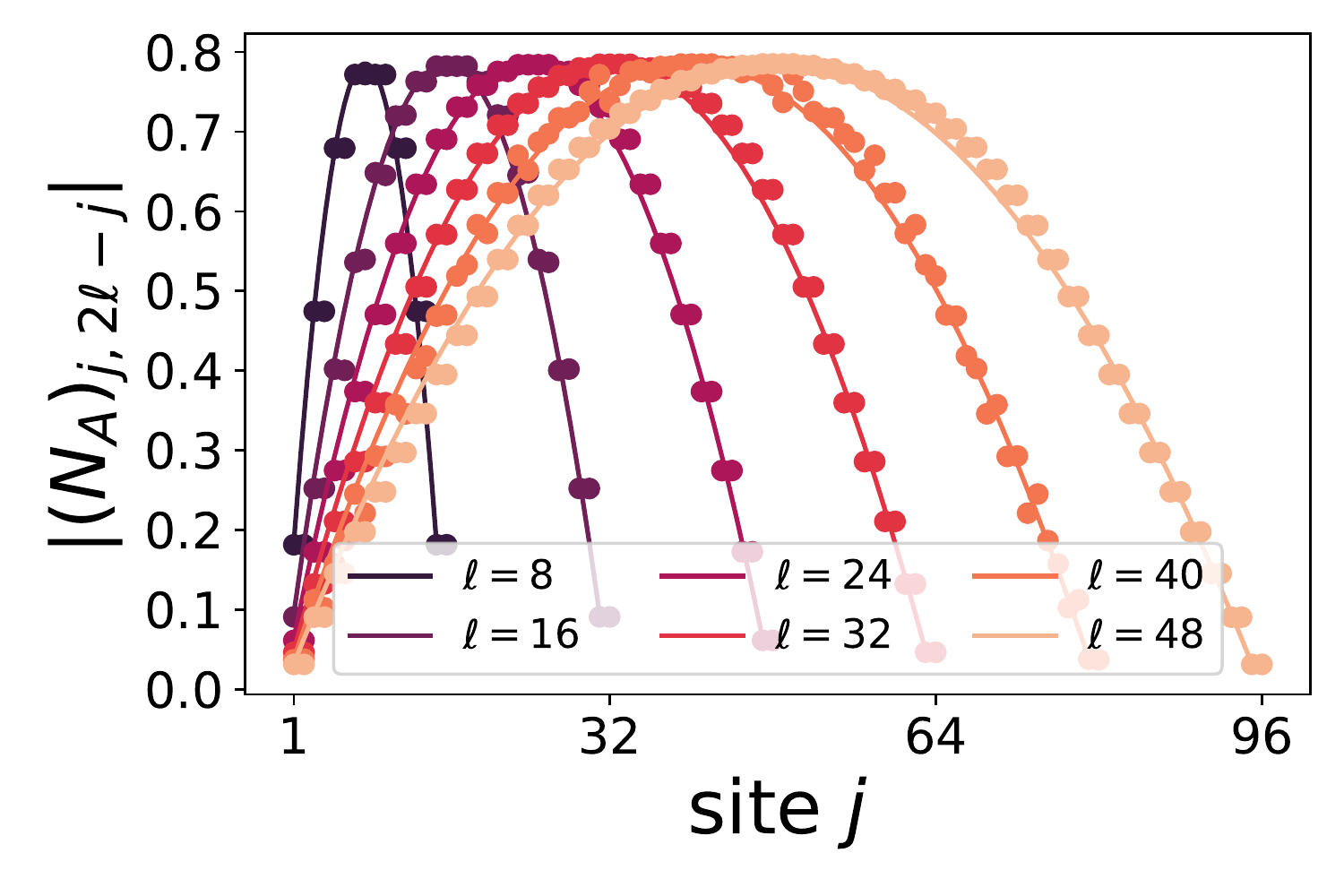}
    \caption{Negativity Hamiltonian corresponding to $\rho_A^{R_1}$ for complex fermions on two adjacent intervals of equal length $\ell$ on the infinite line. The symbols correspond to numerical data, while the solid lines correspond to the discretized form of Eq. \eqref{eq:NH-fcomplex}. The right panel is $N_{A,{\rm loc}}$,  while the left one is $|N_{A,{\rm q-loc}}|$. Using the notation of Eq. (11) of the main text, here we have $a_1=0$, $b_1=\ell=a_2$, $b_2=2\ell$. } 
	\label{fig:is_anti_xx}
\end{figure}

In this section we review the numerical procedure that we used to benchmark our analytical results.
We consider lattice systems described by the quadratic Hamiltonian 
\begin{equation}\label{eq:ff}
    H(\lambda,\gamma)=
    \frac{i}{2}\sum_{l=-\infty}^{\infty}\Big(\frac{1+\gamma}{2}c_{2l}c_{2l+1}\\-\frac{1-\gamma}{2}c_{2l-1}c_{2l+2}+\lambda c_{2l-1}c_{2l} \Big).
\end{equation}
The one-particle energy levels are
\begin{equation}
    \Lambda_{k }=\sqrt{(\lambda-\cos k)^2+\gamma^2\sin^2 k}
\end{equation}
where $k\in [-\pi,\pi]$ is the physical momentum.
For $(\lambda,\gamma)=(1,1)$ or $(\lambda,\gamma)=(0,0)$ the system is critical and Lorentz invariant at low energy. 
In the former case, the critical behavior is described by the conformal field theory of a free massless real fermion with central charge equal to $1/2$ (Majorana) while in the latter case the critical behavior is described by a free massless complex fermion with central charge equal to $1$ (Dirac).
Thus, the Hamiltonian \eqref{eq:ff}, $H(0,0)$ or $H(1,1)$, is the ideal setting to compute the lattice negativity Hamiltonian $\mathcal{N}_A$ and benchmark the analytical expression in Eq. (9) of the main text (real fermion) and in Eq. \eqref{eq:NH-fcomplex} here (complex fermions). 

\begin{figure}
\centering  
\includegraphics[width=0.48\linewidth]{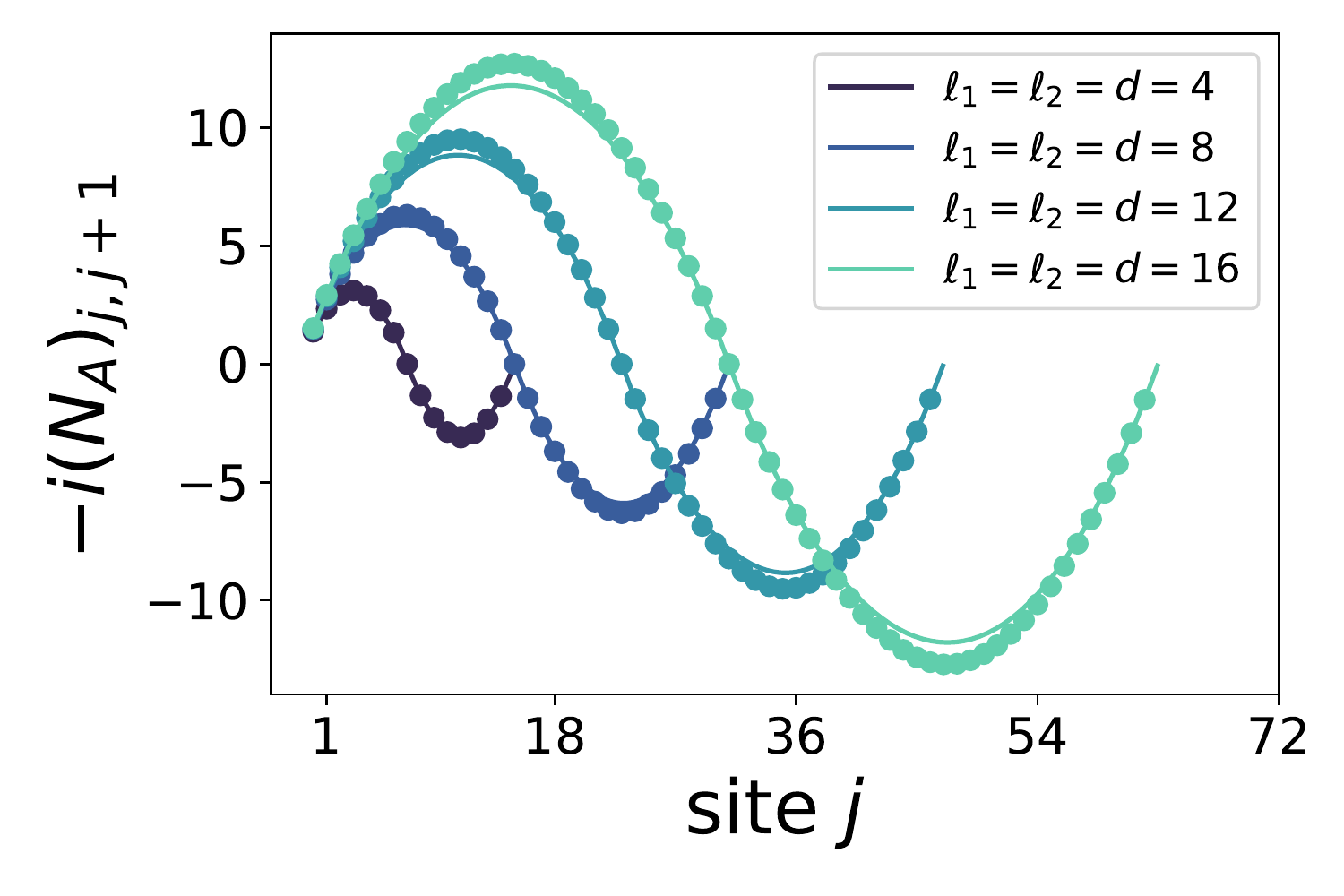}
\includegraphics[width=0.48\linewidth]{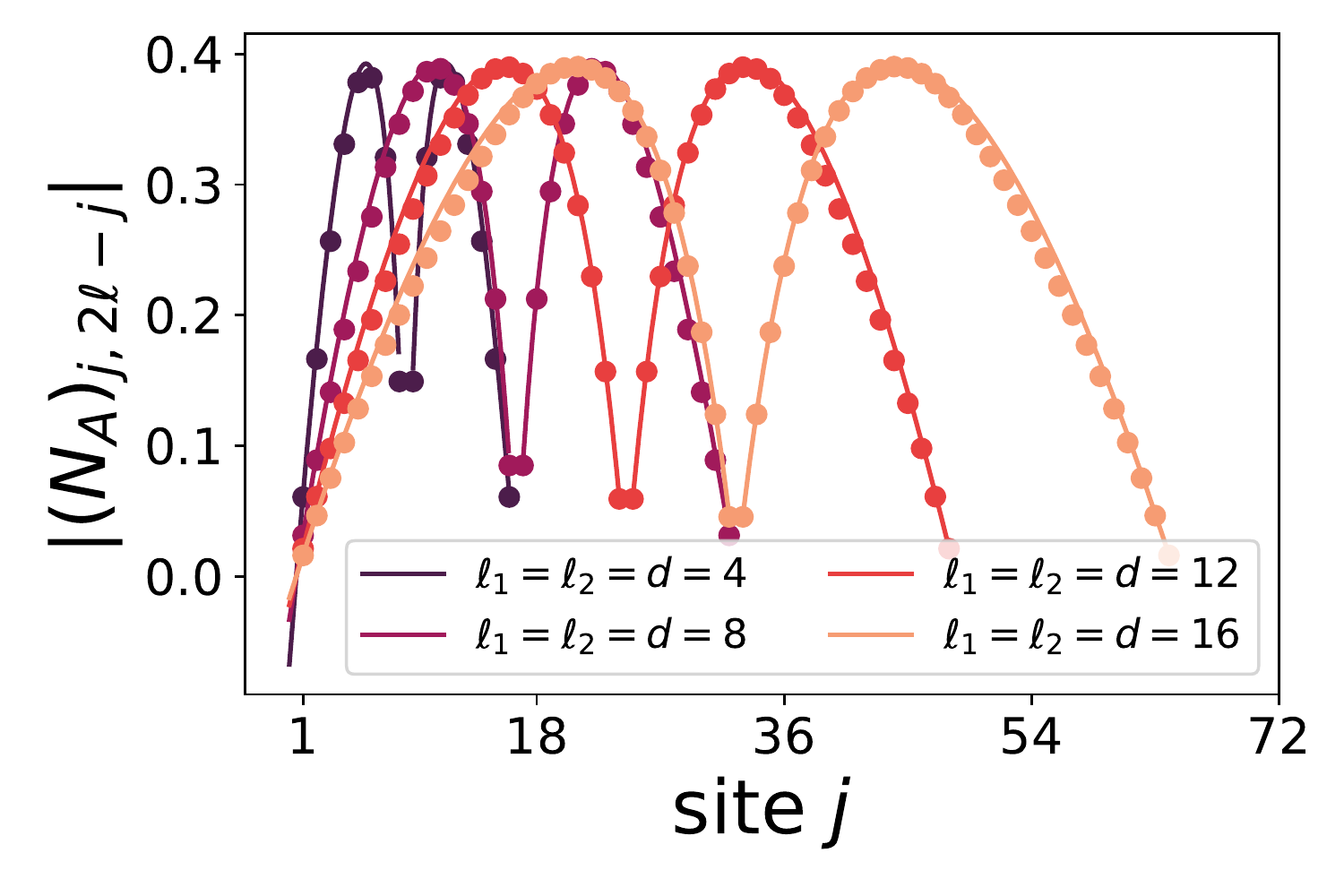}
\caption{Negativity Hamiltonian for a real free fermion and for the geometry of two disjoint intervals of equal length $\ell$ on the infinite line.
The symbols correspond to numerical data, while the solid lines correspond to the discretized form of $N_A$ (Eq. (9) of the main text), both the local part (left panel) and the quasi-local one (right panel).} 
	\label{fig:supp_mat1}
\end{figure}

Let us now consider the ground state of the Hamiltonian \eqref{eq:ff}.
For free complex fermions ($H(0,0)$), the covariance matrix $\Gamma$ is given as
\begin{equation}
    \Gamma_{2j_1-1,2j_2}=-\Gamma_{2j_2-1,2j_1}=i(2C_{j_1,j_2}-\delta_{j_1,j_2}),
\end{equation}
with $C_{ij}=f_{j-i}$  and $f_j$ 
\begin{equation}\label{eq:fj}
    \begin{aligned}
    &f_j=\frac{1}{\pi j} \sin{\frac{\pi j}{2}}, \quad f_0=\frac{1}{2}.
\end{aligned}
\end{equation}
For real fermions ($H(1,1$), the elements of the covariance matrix are instead
\begin{equation}\label{eq:ising_corr}
   \Gamma_{2j_1-1,2j_2}=-\Gamma_{2j_2,2j_1-1}=g_{j_2-j_1},
\end{equation}
where 
\begin{equation}
    g_j=-\frac{i}{\pi}\frac{1}{j+\frac{1}{2}}.
\end{equation}
If now we focus on two intervals
$A=A_1 \cup A_2$, adjacent or disjoint and of arbitrary lengths, the correlation matrix $\Gamma_A$ is obtained from the $\Gamma$ above simply restricting to the subsystem of interest and leading to the block structure of Eq. \eqref{Gammablock}.
If the total length of $A$ is $\ell_1+\ell_2$, the covariance matrix has dimension  $2(\ell_1+\ell_2)\times 2(\ell_1+\ell_2)$.
From this, the covariance matrix $\Gamma^+$ corresponding to the fermionic partial transpose is obtained by building Eq. \eqref{eq:gammap}.
As a consequence, the numerical evaluation of the single particle negativity Hamiltonian corresponding to $\rho_A^{R_1}$, which is a Gaussian operator, just amounts to compute $N_A=\log \frac{1+\Gamma_+}{1-\Gamma_+}$.
The case of two adjacent intervals of equal length for the real fermion has been reported in the main text finding small discrepancies up to a few percent between field theory and numerics. Here we substantiate our findings by displaying further tests of our predictions. %

\subsection{Intervals of equal length $\ell_1=\ell_2$}
We start from Fig. \ref{fig:is_anti_xx} reporting the case of two equal adjacent intervals for a complex fermion ($H(0,0)$). As in the main text, the agreement between numerics and field theory is remarkable.
There are small deviations (up to $\sim 6\%$) between the theoretical curves and the numerical computation that however, as also motivated in the main text, they do not affect the lower part of the negativity spectrum and hence any universal aspect of the negativity Hamiltonian, as also found for the entanglement Hamiltonian, see e.g. \cite{EH-l2,EH-l3}. 

We observe that the data in the right panel of Fig. \ref{fig:is_anti_xx} show also some parity (in $\ell$) effects that were not present for the real fermions. Such oscillations are well known finite $\ell$ effects \cite{ce-10,ccn-10} and disappear as $\ell\to\infty$.

We now move to another geometry starting from real fermions. In Fig. \ref{fig:supp_mat1} we report the case of two equal disjoint intervals at distance $d$ and we benchmark once again our analytical result found in Eq. 9 of the main text. The curves again show a good agreement with the numerical computation, since the discrepancy is at most $\sim 6\%$.

\subsection{Intervals of different length $\ell_1\neq \ell_2$}

\begin{figure}
\includegraphics[width=0.48\textwidth]{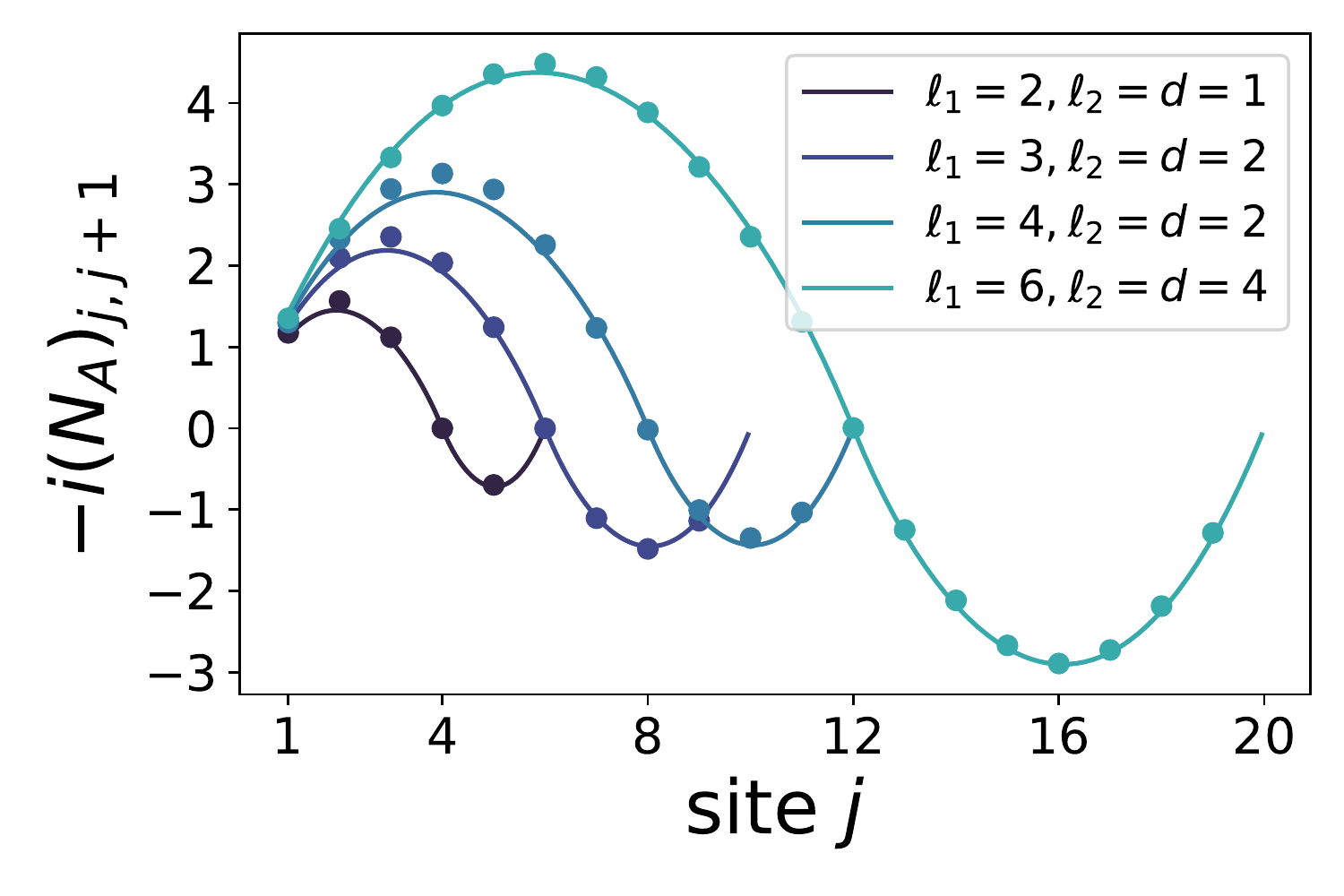}
\includegraphics[width=0.48\textwidth]{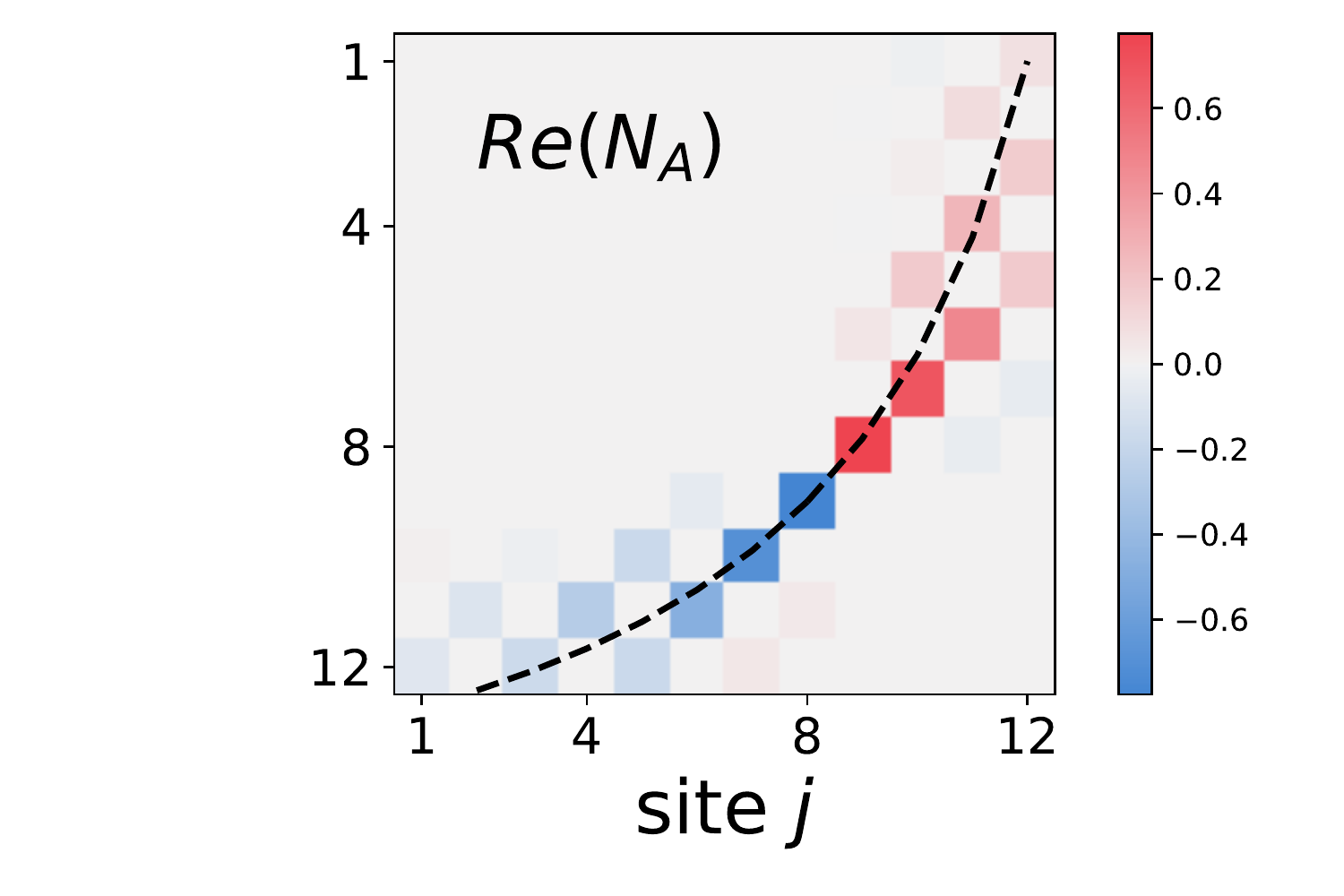}
\caption{Negativity Hamiltonian for a real free fermion and for two disjoint intervals of different length $\ell_1 \neq \ell_2$. 
In this case, the discretized form of $N_A$ correctly reproduces the local behavior of the negativity Hamiltonian (left panel). However, the reflected point $\bar{x}^R$ is not an integer living on the antidiagonal, as in the case of two intervals of equal length. Therefore, we only plot the location of $\bar{x}^R$, Eq. (11) main text, in order to show that its shape is compatible with the structure of the quasi-local part of $N_A$. Here we fix $\ell_1=2\ell_2=2d=4$ (right panel).}
\label{fig:supp_mat2}
\end{figure}

Finally, we analyze in Fig. \ref{fig:supp_mat2} the case of two disjoint intervals of different length, $\ell_1 \neq \ell_2$ for a real fermion. 
In this case, the reflected point $\bar{x}^R$ (Eq. (11) main text) is not on the antidiagonal and does not correspond to an integer number.
Consequently its contributions ``spreads'' to the neighbouring integer. Such an effect is well shown in the right panel of \ref{fig:supp_mat2} in which it is clear that the largest terms of the quasi-local parts of the negativity Hamiltonian are centered around $\bar{x}^R$. 
A more quantitative analysis of the quasilocal terms would require a weighted sum of the nearby elements to get the correct continuum limit, a procedure similar to the one exploited for the entanglement Hamiltonian in Refs. \cite{Arias2017,EH-4,EH-1,EH-2,EH-3}. 
Such analysis is beyond the scope of this work and for this reason we focus on the local term which instead is easily discretized. This is shown in the left panel of Fig. \ref{fig:supp_mat2}. 
Also in this case, the field theory prediction correctly matches the numerics, with small deviations that are at most $\sim 6 \%$ for the system sizes considered.


\end{document}